\documentclass[10pt]{article} 

\usepackage{amsmath}
\usepackage{amssymb}

\usepackage{graphicx}
\usepackage{cite}

\usepackage{color} 

\usepackage[labelfont=bf,labelsep=period,justification=raggedright]{caption}

\makeatletter
\renewcommand{\@biblabel}[1]{\quad#1.}
\makeatother

\newcommand{\ds}[1]{\displaystyle}

\date{}

\begin{document}

\begin{flushleft}
{\Large
\textbf{Senescent fibroblasts can drive melanoma initiation and progression. }
}
\\ %
Eunjung Kim$^{1}$, 
Vito Rebecca$^{2}$,
Inna V. Fedorenko$^{2}$,
Jane L. Messina$^{3}$,
Rahel Mathew$^{3}$,
Silvya S. Maria-Engler$^{4}$,
David Basanta$^{1}$, 
Keiran S.M. Smalley$^{2}$,
Alexander R.A. Anderson$^{1}.$
\\
\bf{1} Integrated Mathematical Oncology Department, H. Lee Moffitt Cancer Center \& Research Institute, Tampa, FL, USA. 
\\
\bf{2} Molecular Oncology, H. Lee Moffitt Cancer Center \& Research Institute, Tampa, FL, USA.
\\
\bf{3} College of Medicine Pathology and Cell Biology, University of South Florida, Tampa, FL, USA.
\\
\bf{4} Department of Clinical Chemistry \& Toxicology, School of Pharmaceutical Sciences, University of S$\tilde{\bf{a}}$o Paulo, S$\tilde{\bf{a}}$o Paulo, Brazil.
\\
$\ast$ E-mail: Eunjung.Kim@moffitt.org
\end{flushleft}

\section*{Abstract}

Skin is one of the largest human organ systems whose primary purpose is the protection of deeper tissues. As such, the skin must maintain a homeostatic balance in the face of many microenvironmental and genetic perturbations. At its simplest, skin homeostasis is maintained by the balance between skin cell growth and death such that skin architecture is preserved. This study presents a hybrid multiscale mathematical model of normal skin (vSkin). The model focuses on key cellular and microenvironmental variables that regulate homeostatic interactions among keratinocytes, melanocytes and fibroblasts, key components of the skin. The model recapitulates normal skin structure, and is robust enough to withstand physical as well as biochemical perturbations. Furthermore, the vSkin model revealed the important role of the skin microenvironment in melanoma initiation and progression. Our experiments showed that dermal fibroblasts, which are an important source of growth factors in the skin, adopt a phenotype that facilitates cancer cell growth and invasion when they become senescent. Based on these experimental results, we incorporated senescent fibroblasts into vSkin model and showed that senescent fibroblasts transform the skin microenvironment and subsequently change the skin architecture by enhancing the growth and invasion of normal melanocytes as well as early stage melanoma cells. These predictions are consistent with our experimental results as well as clinical observations. Our co-culture experiments show that the senescent fibroblasts promote the growth and invasion of non-tumorigenic melanoma cells. We also observed increased proteolytic activity in stromal fields adjacent to melanoma lesions in human histology. This leads us to the conclusion that, senescent fibroblasts create a pro-oncogenic environment that synergizes with mutations to drive melanoma initiation and progression and should therefore be considered as a potential future therapeutic target. This study suggests a potential link between aging in the skin microenvironment and the development of melanocytic neoplasms.

\section*{Author Summary}

Skin homeostasis depends upon the complex interplay of skin cells as well as interactions between cells and the microenvironment. Here, we generated a virtual skin (vSkin) model in order to test our hypothesis that dysregulation of cell-microenvironment interaction leads to aberrant skin structure and furthermore recapitulates pathologic conditions of the skin. To this end, we used the hybrid cellular automata method. The model couples a cellular automata that describes biological rules for cell types with partial differential equations that describe continuous microenvironmental factors. Our simulation recapitulated normal skin growth as well as self-repair. It also showed that microenvironmental factors produced by stromal cells (fibroblasts) play an important role in maintaining normal skin homeostasis. We find that the transdifferentiation of fibroblasts due to senescence changes the skin microenvironment to drive melanoma initiation and progression. Our model gives new insights into the processes of melanoma initiation and progression, and suggests a novel strategy for treatment.

\section*{Introduction}
Skin is the largest organ of the body. It is a physical barrier that limits the flow of water and electrolytes, while providing protection from ultraviolet radiation, microorganisms and toxic substances. Human skin consists of three layers, the epidermis, dermis, and subcuits. The epidermis consists of keratinocytes, melanocytes, Langerhans cells and Merkel cells. The dermis is composed of connective tissue, fibroblasts, blood vessels and lymphatics. The subcutis is made up of loose connective tissue and insulating fat cells. Keratinocytes are the main constituent of the epidermis, composing around 95\% of the total cell number \cite{Rook:2010fk}. Within the epidermis the keratinocytes form a stratified squamous epithelium in which the keratinocytes are tightly connected to each other through desmosomes. Like other epithelial tissue, the epidermis continuously renews itself. In the epidermis, cell proliferation occurs at the basal layer followed by a stepwise upward migration and maturation of the cells. As the keratinocytes mature they lose their nuclei and turn into keratin-filled corneocytes that form a 10-30 cell thick water resistant protective layer that gives the skin its critical barrier function. Protection of the keratinocytes from the DNA-damaging effects of ultraviolet radiation is mediated by the pigment producing cells, melanocytes, which locate to the basement membrane of the skin in a 1:5 ratio with the basal keratinocytes~\cite{FITZPATRICK:1963fk,Haass:2005p106}. Upon exposure to UV radiation, melanocytes produce melanin which is transferred to the surrounding keratinocytes via an active transport process. Once taken up into keratinocytes, melanosomes orientate themselves over the nucleus in a protective ``hat" that shields the nuclei of the basal keratinocytes from UV-induced DNA damage~\cite{Costin:2007p2465}. Keratinocytes in turn regulate the growth and behavior of melanocytes through a complex network of cell-cell adhesion proteins and secretory factors~\cite{Haass:2005p106}. In addition to these interactions, the structure and strength of skin is also critically dependent upon the extracellular matrix (ECM) and the fibroblasts in the dermis. Dermal fibroblasts secrete ECM components and are responsible for mechanical strength of the dermis~\cite{Sorrell:2004bh} as well as producing growth factors such as epidermal growth factors (EGF), transforming growth factor $\beta$ ($\mbox{TGF}_\beta$), and basic fibroblast growth factor (bFGF)~\cite{Kalluri:2006uq}. These growth factors promote growth and facilitate the constant renewal of the epidermis.

Melanoma is the most devastating form of skin cancer~\cite{Chin:2003uq,Miller:2006fk}, arising from the malignant transformation of melanocytes. The first phenotypic change in melanocytes is the disruption of growth controls and development of melanocytic nevi, a benign mole. At the molecular level, the growth is stimulated by the abnormal activation of the mitogen-activated protein kinase (MAPK) signaling pathway which can arise through activating mutations in the {\it NRAS} (15\% of melanomas) or {\it BRAF} (50\% of melanomas)~\cite{Albino:1989kx,Davies:2002fk,Omholt:2003vn,Houben:2008uq}. Despite {\it BRAF} mutations being important for melanoma development, the majority of benign nevi also harbor {\it BRAF} mutations~\cite{Pollock:2003ys} and rarely undergo full malignant transformation. Instead, benign nevi cease proliferation and remain quiescent for decades, entering a state of oncogene-induced senescence (OIS)~\cite{Michaloglou:2005p9307,Dhomen:2009vn}. This suggests that nevi must acquire additional intrinsic (molecular) or extrinsic (microenvironmental) damages to free themselves from the OIS state and develop into full-blown malignant melanoma. There is some suggestion that activation of phosphoinositide 3-kinase (PI3K)/protein kinase B (PKB or AKT) due to loss of a tumor-suppressor gene, phosphatase and tensin homologue (PTEN) may be required for escape from senescence~\cite{Vredeveld:2012ys}. The suggested mechanism is that the inactivation of PTEN fails to attenuate the PI3K/AKT level, and consequently increases PI3K/AKT expression that facilitates cell proliferation and survival. However, the precise molecular events that underlie the transformation of  nevus cells develop into melanoma are not well understood.

Cancer is typically a disease of old-age, and there is increasing evidence that senescence within the stroma, particularly in the fibroblast compartment can drive tumor development. A number of reports have shown that senescent stromal fibroblasts stimulate premalignant and malignant epithelial cells to grow in cell culture and to form tumors in mice~\cite{Krtolica:2001zr,Krtolica:2002ly,Bhowmick:2004p7596,Parrinello:2005ve}. Mechanistically, this seems to involve the secretion of factors from senescent fibroblasts, such as Matrix metalloproteinase-3 (MMP-3) and Interleukin-6 (IL-6), that in turn remodel the microenvironment, alter epithelial differentiation, promote endothelial cell motility and stimulate cancer cell growth both {\it{in vitro}} and {\it{in vivo}}~\cite{Coppe:2008p7159,Laberge:2012kh}. It is well known that aged skin tends to harbor large numbers of senescent fibroblasts~\cite{Campisi:1998fu}, and there is a direct correlation between age and the expression of $\mbox{p16}^{INK4A}$ (a cell cycle inhibitor) expression within the dermis~\cite{Ressler:2006zr}. In other words, concentrations of $\mbox{p16}^{INK4A}$ increase dramatically as tissue ages. Photo-damage following exposure to UV radiation can also increase the level of senescence within the skin through the effects of oxidative damage upon the telomeres shortening~\cite{Zglinicki:2000qc,Zglinicki:2002ss}. Although the exact role of the dermal microenvironment in melanoma development has been little studied, there is evidence that the mildly hypoxic environment of the skin contributes to melanocyte transformation~\cite{Bedogni:2005mw}. It is also known that fibroblasts contribute towards the survival of early-stage melanoma cells by secreting the growth factor IGF-1 and that stromal remodeling through $\mbox{TGF}_\beta$ overexpression can enhance melanoma survival in mice~\cite{Li:2003p77}.

In order to gain a better insight into the mechanisms underlying normal skin homeostasis, and ultimately the role of the stroma in melanoma initiation and progression, we developed a virtual skin model. Several computational studies of skin have recently been developed. Deterministic ordinary differential equations have been used to model immune signaling in human skin, providing a quantitative strategy that distinguishes healthy from pathologic inflammatory responses~\cite{Valeyev:2010fk}. A similar method has been employed  to model mouse embryonic melanoblast proliferation dynamics~\cite{Aylaj:2011uq}. Another continuous modeling framework, a system of deterministic reaction-diffusion equations, has been applied to model tumor-immune interactions~\cite{Eikenberry:2009p2090}. A continuum fluid mechanics approach has been adopted to analyze the stability of the interface between epithelia and stroma~\cite{Basan:2011vn}. A mixture theory has been used for structure stability analysis of melanoma growth\cite{Ciarletta:2011kx} and microstructure patterning~\cite{Chatelain:2011ys}. More discrete based modeling frameworks have also been applied to model skin. Agent based methods have been used to model epidermal keratinocyte behavior in normal skin~\cite{Grabe:2005p8047,Suetterlin:2009ys, Adra:2010p3575}, epidermal homeostasis control~\cite{Schaller:2007uq}, keratinocyte originated skin disease~\cite{Grabe:2007p7951}, the interactions between keratinocytes and fibroblasts~\cite{Sun:2008p3760}, and melanocyte distribution in epidermis~\cite{Thingnes:2012fk}. However, none of these previous studies have considered normal skin homeostasis and its disruption as a route to melanoma development.
 
 In this study, we have developed a hybrid multiscale mathematical model that simulates normal skin homeostasis. We named the model virtual skin (vSkin). We employed a hybrid cellular automata (HCA)~approach \cite{anderson1997, Anderson:1998fk, AndersonInAlt:2003p11378,Anderson:2005p10258}. The general modeling approach is to derive a set of rules for discrete cells and couple this discrete cellular population with a suite of continuous microenvironmental factors. Previously, we have applied this method to investigate tumor-stroma interactions in prostate cancer progression~\cite{Basanta:2009p10235}. In this study, we developed a minimal set of cell rules for melanocytes, keratinocytes, and fibroblasts, which determine each cell's life cycle. Then, we developed a cell interaction network that defines local interactions between cells and their microenvironment. These local interactions lead to the emergence of normal skin structure that recapitulates the {\it in vivo} structure of skin. Using our model, we observed that the vSkin model is robust enough to withstand perturbations. The vSkin model not only recovers from massive loss of its constituents but also finds an equilibrium state after super-physiological perturbations of microenvironmental factors. The proposed model also reveals the possible involvement of these microenvironmental factors in conjunction with senescent stroma in driving melanoma initiation and progression. 
 
The paper is organized as follows. In the first section, we give an overview of the vSkin model development process. In the second section, we initially present results showing the stability of vSkin. Then, we discuss the robustness of the vSkin as well as application of the model to melanoma initiation via melanocyte mutation and fibroblast transdifferentiation. We conclude with a discussion focusing on the implications of the model predictions in terms of a novel strategy for treatment.
 
 \begin{figure}[h]
\begin{center}
\includegraphics[width=4in]{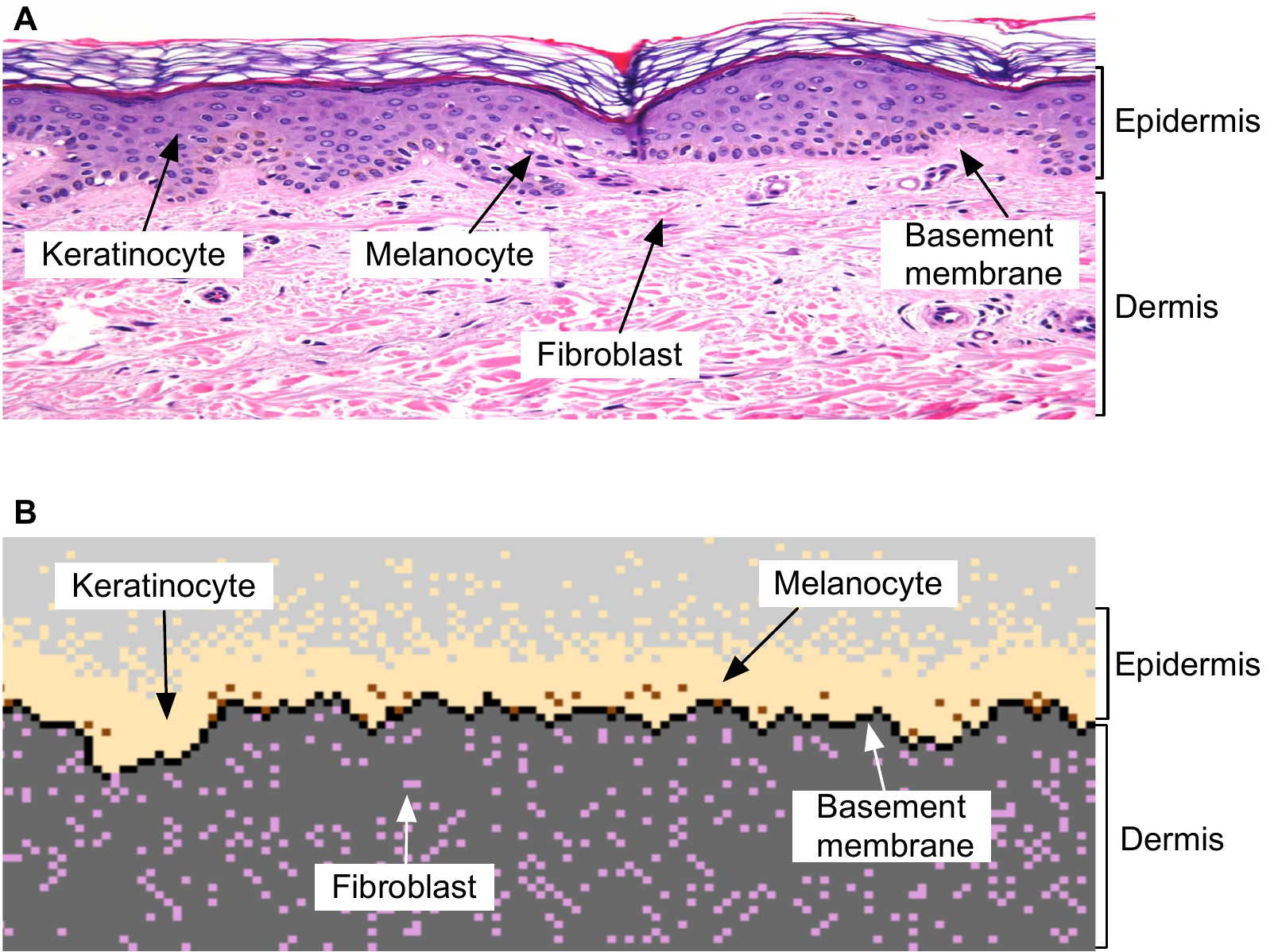}
\end{center}
\caption{
{\bf Human skin normal structure.} A: H $\&$ E stained cross-section of human normal skin at 20X showing the epidermis contains a basal layer, melanocytes and keratinocytes. The dermis contains  fibroblasts and extracellular matrix. B: Model domain with its key cell types such as keratinocyte, melanocyte and fibroblasts. The density of extracellular matrix is considered and assumed to be continuous in the domain. The gray  color represents the density of extracellular matrix at each node in the domain, i.e., darker gray represents denser matrix. The basement membrane separates epidermis and dermis layer in the model, and it is represented by continuously connected nodes where the density of matrix has reached its maximum (1.0).
 }
\label{Fig1}
\end{figure}
  
\section*{Material and methods} 

\subsection*{Mathematical model}
We propose a systemic model of skin capable of growth control as well as self-repair.  In this section, we describe the development of the vSkin model, which represents a cross-section of skin as shown in Figure \ref{Fig1}. We first present key microenvironmental variables involved in normal skin structure and function. Then we move on to discuss the implementation of key cell phenotypes.

\subsubsection*{Hybrid cellular automata model}

\begin{figure}[h]
\begin{center}
\includegraphics[width=4in]{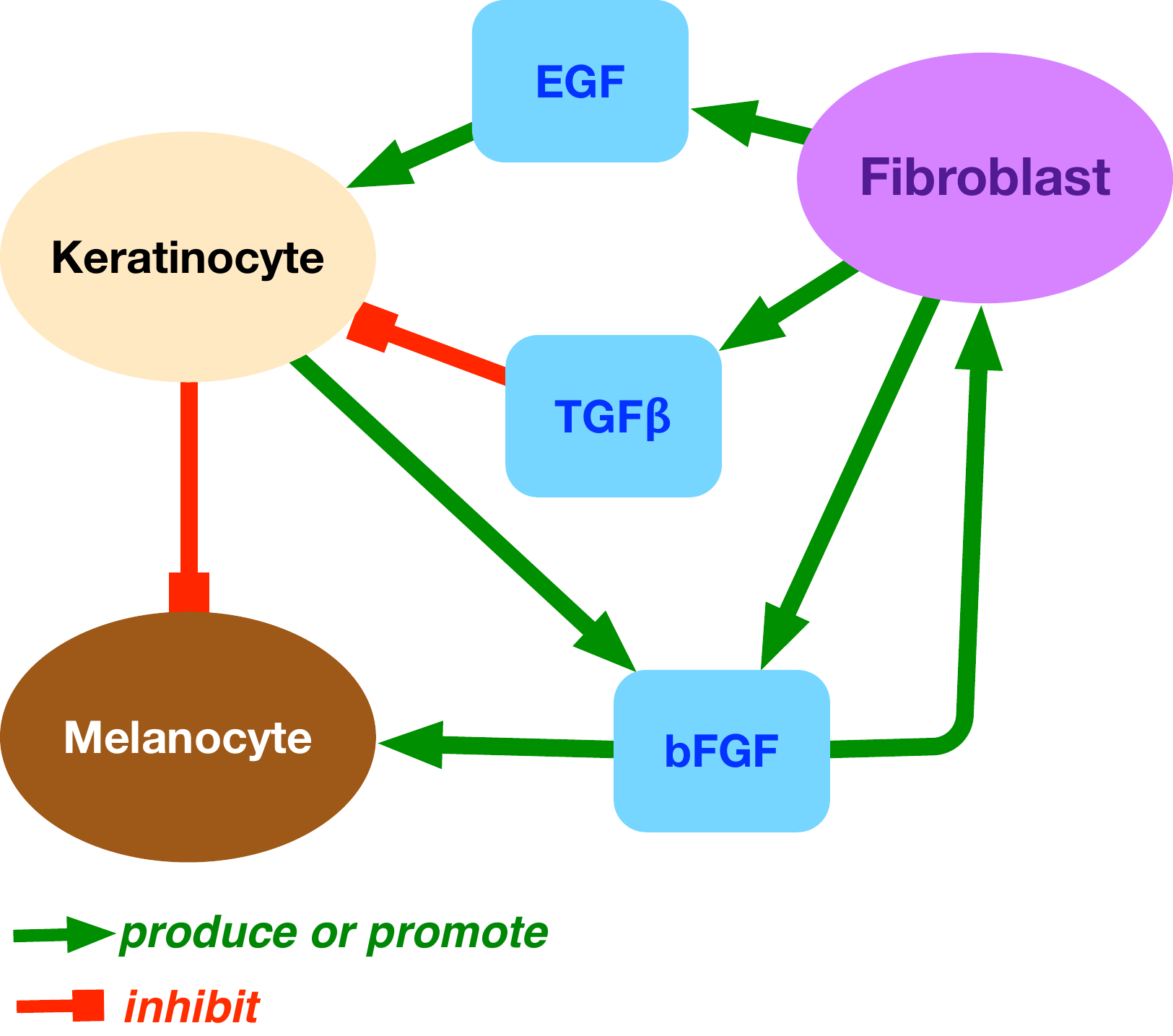}
\end{center}
\caption{
{\bf Cell interaction network for normal skin.} Green lines represent promoting events, while red lines imply inhibiting effects. Note, cell coloration is used in all simulations to represent the equivalent cell. In the model, a fibroblast is assumed to produce three growth factors, Epidermal growth factor (EGF), Basic fibroblast growth factor (bFGF) and Transforming growth factor ($\mbox{TGF}_\beta$). EGF promotes growth of keratinocytes, whilst $\mbox{TGF}_\beta$ inhibits the growth. The growth of melanocyte is controlled by both bFGF and physical interaction (contact inhibition) with neighboring keratinocytes.}
\label{Fig2}
\end{figure}

 In order to keep the complexity of the model to a minimum, we only consider three key cell types critical for skin function, melanocytes, keratinocytes, and fibroblasts. These cells are defined as points on a two-dimensional grid that represents a cross-section of human skin ($6~\mbox{mm} \times 1.24~\mbox{mm}$), as shown in Figure \ref{Fig1}B.  Each grid point may also be occupied by up to five microenvironmental variables, EGF, bFGF, $\mbox{TGF}_\beta$, MMPs, and  basement membrane/ECM. The discrete cells are coupled with these microenvironmental concentrations to define the hybrid cellular automata (HCA) model. The cells on the lattice influence these microenvironmental concentration fields through production and consumption, but are in turn affected by the concentrations, as they serve as regulators of cell behavior.
To better understand how these cells interact with one another and their microenvironmental variables we developed a cell interaction network (Figure \ref{Fig2}). This network shows the key interactions that are believed to be important in maintaining normal skin structure and function. The network was derived through lengthy discussion with our experimental co-authors and an extensive literature review, detailed throughout the following sections. 

\subsubsection*{Microenvironmental factors}

Keratinocytes, melanocytes and fibroblasts require a large number of different growth factors and adhesion signals to promote and maintain their normal growth and maintenance. However, here we restrict ourselves to either one or two chemical variables for each cell type in the model for the sake of simplicity. For a keratinocyte, we consider EGF and $\mbox{TGF}_\beta$, because these two factors are known to be the most effective regulators for keratinocyte growth control~\cite{Green:1985vn,Nanney:1986ys,Krane:1991zr, Hannon:1994fk,Ewen:1996uq}.  We consider bFGF for the growth control of both melanocytes~\cite{Shih:1993zr} and fibroblasts~\cite{Makino:2010ly}. Since fibroblasts produce protease when they are activated~\cite{Sorrell:2004bh,Kalluri:2006uq}, we include MMP in the model as well. Finally, ECM concentration is incorporated, since ECM is required for structural support, and it separates the dermal and epidermal layers~\cite{Rook:2010fk,Sorrell:2004bh} in the form of basement membrane. The dynamics of the five microenvironmental variables are defined by a system of partial differential equations that describe how each of them evolves in space and time. Since these continuous variables interact with discrete cells, for clarity we present only the discretized version of each equation.

The discretized governing equation for EGF ($G_E$) at a node ${\eta} \equiv (\eta_x,\eta_y) \in \Omega \subset \mathbb{Z}^2$ is 
\begin{equation}
 \frac{[G_E]^{+1}-[G_E]}{\delta t}  = \overbrace{D_E \Delta_2 [G_E]}^{\mbox{\tiny {diffusion}}} +  \overbrace{(\alpha_E F_\eta+\beta_E K_{\eta})}^{ \begin{array}{c} \mbox{\tiny{production by}}\\ [-0.25in] \\ \mbox{\tiny{ fibroblasts and keratinocytes}}\end{array}} - \overbrace{(\gamma_E K_{\eta}[G_E]+\zeta_E M_{\eta}[G_E])}^{\begin{array}{c}\mbox{\tiny{binding to EGF receptors of}} \\ [-0.25in]\\ \mbox{\tiny{keratinocytes or melanocytes}}\end{array}} - \overbrace{\lambda [G_E]}^{\mbox{\tiny{decay}}},
 \label{eq1}
\end{equation}
where
\begin{equation*}
X_\eta =
 \begin{cases}
1 & \mbox{if a node } { \eta} \mbox{ is occupied by a cell type } X ~\mbox{at time step}~t, \\ 0 & \mbox{otherwise,}
\end{cases}
 \label{cell}
\end{equation*}
where a cell type $X$ can be keratinocyte ($K$), melanocyte ($M$) or fibroblast ($F$).
In Eq (\ref{eq1}), $[\cdot]$ represents the concentration of a chemical at a node ${\eta}$ at time step $t$ (e.g.,$[G_E] = G_E(t,{\eta_x,\eta_y}))$, $[\cdot]^{+1}$ represents the concentration at the next time step, and $\delta t$ represents a time step. The operator $\Delta_2$ defines a two-dimensional discrete Laplacian,
\begin{equation*}
\Delta_2 f (t, \eta_x,\eta_y) \equiv \frac{f(t,\eta_x+\delta h, \eta_y) + f(t,\eta_x - \delta h, \eta_y) + f(t,\eta_x,\eta_y+\delta h) + f(t,\eta_x,\eta_y-\delta h) - 4f(t,\eta_x,\eta_y) }{(\delta h)^2}, 
\end{equation*}
where $\delta h$ is the grid size.
Equation (\ref{eq1}) defines EGF as a diffusing chemical which is increased by a rate of $\alpha_E$ or $\beta_E$ when a grid node $\eta$ is occupied by a fibroblast ($F$) or a keratinocyte ($K$), respectively.  The concentration is decreased if a node is occupied by a keratinocyte or melanocyte ($M$) at a rate of $\gamma_E$ and $\zeta_E$, respectively, due to uptake of EGF. Lastly, the concentration has a natural decay rate of $\lambda$ per time step. Similarly, the dynamics of bFGF ($G_b$) is modeled as follows,
\begin{equation}
\begin{array}{cc}
&\displaystyle{\frac{[G_b]^{+1}-[G_b]}{\delta t}}  =  \overbrace{D_b \Delta_2 [G_b]}^{\mbox{\tiny {diffusion}}} +   \overbrace{(\alpha_b F_\eta + \beta_b K_\eta+\alpha_b^s F_\eta^s + \beta_b^s M_\eta^s)}^{ \begin{array}{c} \mbox{\tiny{production by  fibroblasts, keratinocytes,}}\\ [-0.25in] \\ \mbox{\tiny{ abnormal fibroblasts and transformed melanocytes}} \end{array}} \\
 & - \overbrace{(\gamma_b K_\eta [G_b]+\zeta_b M_\eta [G_b]+\xi_b F_\eta [G_b] +\zeta_b^s M_\eta^s [G_b]+\xi_b^s F_\eta^s [G_b])}^{\begin{array}{c}\mbox{\tiny{binding to bFGF receptors of keratinocytes,}} \\ [-0.25in]\\ \mbox{\tiny{ melanocytes, fibroblasts, transformed melanocytes, and abnormal fibroblasts}}\end{array}} - \overbrace{\lambda [G_b]}^{\mbox{\tiny{decay}}},
 \end{array}
\label{eq2}
 \end{equation}
where fibroblasts and keratinocytes produce bFGF at the rate of $\alpha_b$ and  $\beta_b$, respectively, and transformed fibroblasts and transformed melanocytes produce bFGF at the rate of $\alpha_b^s$ and  $\beta_b^s$, respectively.  All cell types bind to bFGF at a rate of  $\gamma_b$, $\zeta_b$, $\xi_b$, $\zeta_b^s$, and $\xi_b^s$. 
Finally, $\mbox{TGF}_\beta$ ($T_\beta$) is produced by fibroblasts and keratinocytes at a rate of $\alpha_{T_\beta}$ and $\beta_{T_\beta}$, respectively, diffuses at a rate $D_{T_\beta}$, and binds to keratinocytes, melanocytes (or transformed melanocytes), fibroblasts (or abnormal fibroblasts) at rates of  $\gamma_{T_\beta}$, $\zeta_{T_\beta}$, $\xi_{T_\beta}$, respectively.
 \begin{equation}
 \begin{array}{cc}
& \displaystyle{\frac{[T_\beta]^{+1}-[T_\beta]}{\delta t}}    = \overbrace{D_{\beta} \Delta_2 [T_\beta]}^{\mbox{\tiny {diffusion}}} +  \overbrace{(\alpha_{T_\beta} F_\eta +\beta_{T_\beta} K_\eta)}^{ \begin{array}{c} \mbox{\tiny{ production by}}\\ [-0.25in] \\ \mbox{\tiny{fibroblasts and keratinocytes} }\end{array}} \\
 &  - \overbrace{(\gamma_{T_\beta} K_\eta [T_\beta] +\zeta_{T_\beta} M_\eta [T_\beta]+ \xi_{T_\beta} F_\eta [T_\beta] +\zeta_{T_\beta} M_\eta^s [T_\beta]+ \xi_{T_\beta} F_\eta^s [T_\beta])}^{\begin{array}{c}\mbox{\tiny{binding to $\mbox{TGF}_\beta$ receptors of }} \\ [-0.25in]\\ \mbox{\tiny{keratinocytes, melanocytes, fibroblasts, transformed melanocytes, and abnormal fibroblasts}}\end{array}} - \overbrace{\lambda [T_\beta]}^{\mbox{\tiny{decay}}}.
 \end{array}
\label{eq3}
\end{equation}

ECM (E) is produced by both fibroblasts or keratinocytes depending on the local concentration of ECM. In other words, keratinocytes and fibroblasts produce ECM with a rate of ${\kappa_E}$ and $\rho_E$ only if there is any loss compared to the initial density of epidermis ($E_0$) and dermis ($E_1$), respectively. However, transformed fibroblasts always produce ECM with a rate of $\rho_E^s$. Whenever fibroblasts are in contact with keratinocytes or melanocytes, fibroblasts can make much denser matrix which will create new basement membrane~\cite{Andriani:2003ve}.  Basement membrane (BM) is defined as a continuously connected curve of grid points, each of whose ECM densities is maximal (taken to be 1 in non-dimensional units). ECM is degraded by MMP at a rate of $\mu$. The governing equation for ECM is

\begin{equation}
\frac{[E]^{+1}-[E]}{\delta t}   =   \overbrace{\kappa_{E} {H}(E_0 - [E]) K_\eta}^{ \begin{array}{c} \mbox{\tiny{ production}}\\ [-0.25in] \\ \mbox{\tiny{by keratinocytes} }\end{array}} +  \overbrace{\rho_{E} {H}(E_1 - [E]) F_\eta}^{ \begin{array}{c} \mbox{\tiny{ production}}\\ [-0.25in] \\ \mbox{\tiny{by fibroblasts} }\end{array}} 
 + \overbrace{\rho_{E}^s F_\eta^s}^{ \begin{array}{c} \mbox{\tiny{ production}}\\ [-0.25in] \\ \mbox{\tiny{by abnormal fibroblasts} }\end{array}}  - \overbrace{\mu [P][E]}^{ \begin{array}{c}\mbox {\tiny{degradation}}\\ [-0.25in]\\ \mbox{\tiny{by MMP}}\end{array}} 
\label{eq4}
\end{equation}
where $E_0$ represents initial density of epidermis and $E_1$ is that of dermis. Note that when a fibroblast is in contact with either a melanocyte or a keratinocyte, we set $E_1$ to be its maximal value 1.0 to model fibroblast's basement membrane generation. The function ${H}$ is the Heaviside step function defined as

\begin{equation}
H(x)=
 \begin{cases}
0 & \mbox{if } x \leq 0 \\ 1 & \mbox{if } x>0.
\end{cases}
 \label{stepFunc}
\end{equation}
MMPs (P) are produced by abnormal fibroblasts, diffuse at a rate $D_{p}$, and are depleted as they degrade the ECM (E). Note that when melanocytes are transformed, they can produce MMPs as well. In the model, we assume that transformed melanocytes produce MMPs with a rate of $\beta_p^s$. The governing equation is
 \begin{equation}
 {
\frac{[P]^{+1}-[P]}{\delta t}  =    \overbrace{D_{p} \Delta_2 [P]}^{\mbox{\tiny {diffusion}}} + \overbrace{\alpha_{p}^s  F_\eta^s}^{ \begin{array}{c} \mbox{\tiny{ production}}\\ [-0.25in] \\ \mbox{\tiny{by abnormal fibroblasts} }\end{array}} + \overbrace{\beta_{p}^s  M_\eta^s}^{ \begin{array}{c} \mbox{\tiny{ production}}\\ [-0.25in] \\ \mbox{\tiny{by transformed melanocytes} }\end{array}} - \overbrace{\mu [P][E].}^{ \begin{array}{c}\mbox {\tiny{degradation}}\\ [-0.25in]\\ \mbox{\tiny{by MMP}}\end{array}}}
\label{eq5}
\end{equation}

All boundary conditions for all microenvironmental variables are no-flux on the surface and the bottom of the domain. Periodic boundary conditions were imposed on the left and right sides of the domain.

\begin{figure}[h]
\begin{center}
\includegraphics[width=4in]{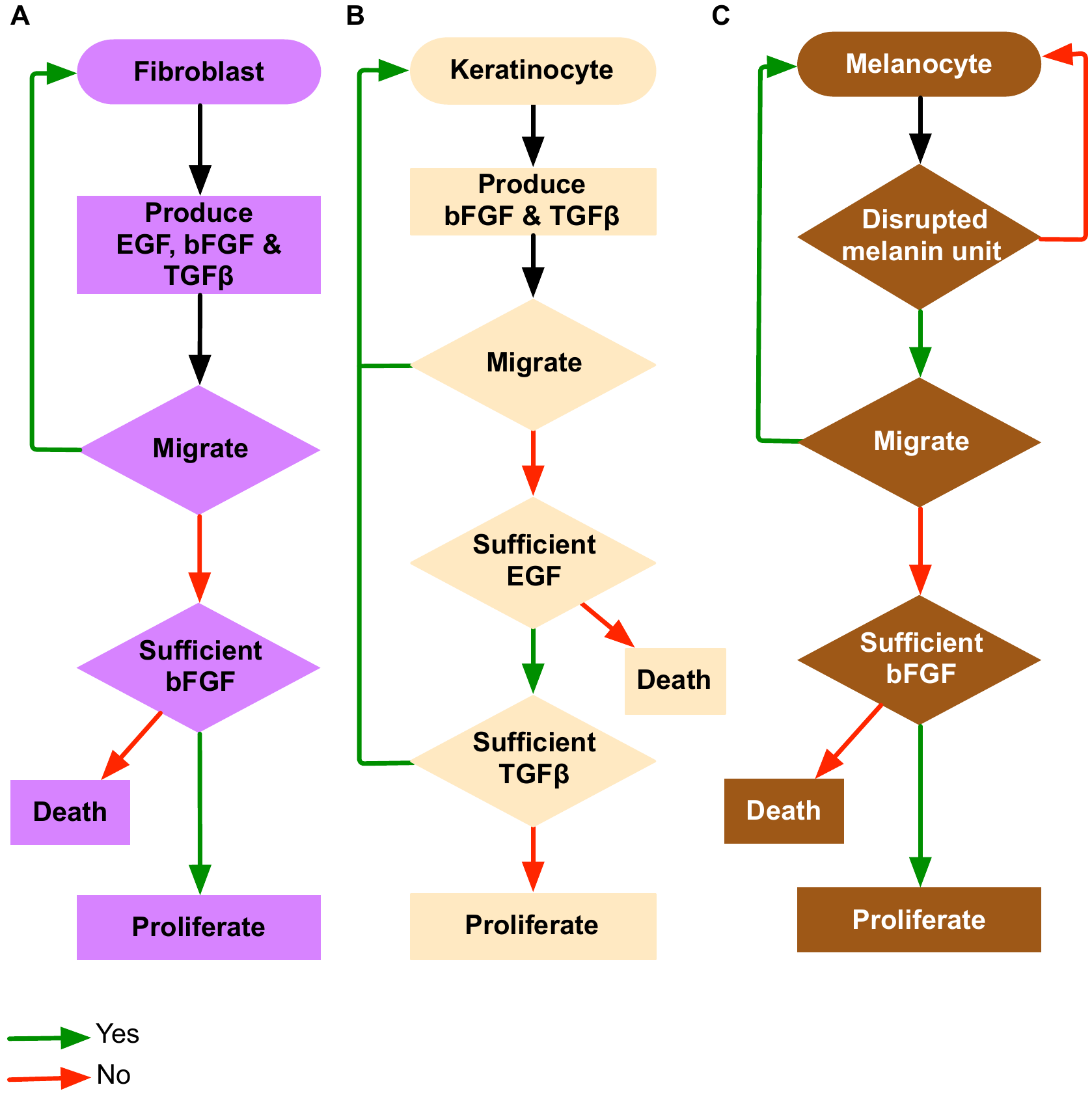}
\end{center}
\caption{
{\bf Flowcharts showing the cell life cycle for fibroblasts, keratinocytes and melanocytes.} A: Fibroblasts produces bFGF, EGF and  $\mbox{TGF}_\beta$, and proliferate if there is enough bFGF. Low concentrations of bFGF trigger fibroblast death. B: Keratinocytes produces bFGF, and proliferate when there is enough EGF and a low concentration of $\mbox{TGF}_\beta$. An insufficient level of EGF triggers keratinocyte death. C: Melanocyte start their life cycle by checking the melanin unit (the number of keratinocyte neighbors) and then decide to migrate or not. Sufficient bFGF stimulates melanocyte proliferation while insufficient bFGF triggers cell death.}
\label{Fig3}
\end{figure}

\subsubsection*{Cell phenotypes}
The behavior of each cell depends on its neighbors and the microenvironmental concentrations in which it resides. In the vSkin model, we consider the von Neumann neighborhood of range 1 (i.e., four orthogonal neighbors) for each grid point $\eta \in \Omega$, denoted by  ${\it{N}}_\eta$. In vSkin, each cell has three possible phenotypes: proliferative, migratory and dead.  In this section, we describe how these phenotypes are implemented. Biological rules for each cell were abstracted from a literature review and are summarized as flowcharts in Figure~\ref{Fig3}. For the sake of simplicity, each cell is allowed to execute only one phenotype, proliferative, migratory, or dead, at each cell time step ($t_c$). In other words, if the cell chooses one phenotype at current time step, it exits current life cycle and waits until next time step. If a cell does not execute any of these three phenotypes, it remains as a quiescent cell.

\begin{figure}[!ht]
\begin{center}
\includegraphics[width=4in]{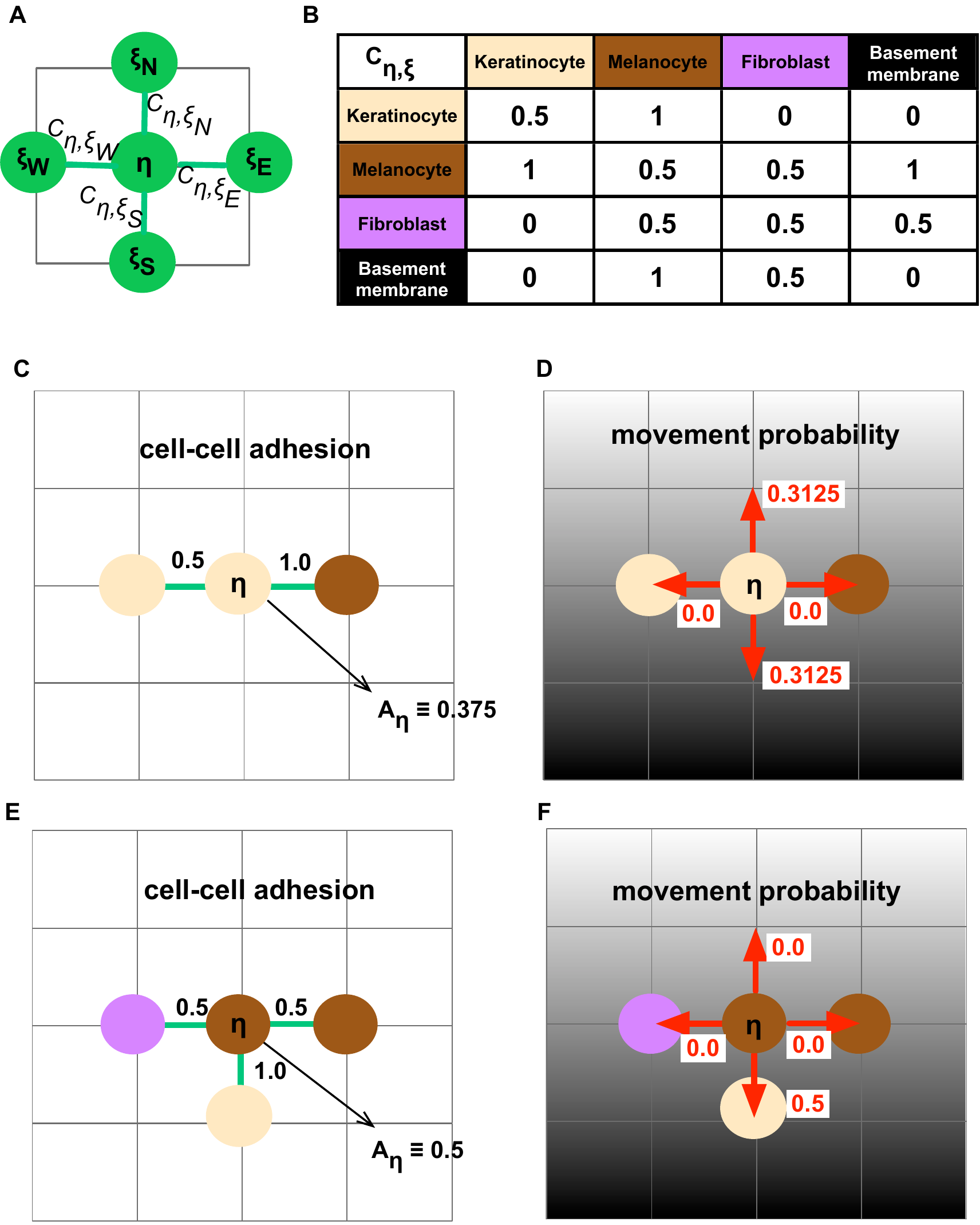}
\end{center}
\caption{
{\bf Cell-cell adhesion and migration} A:  The cell-cell adhesion value $A_\eta$ at each node is an average of its four orthogonal cell-cell adhesive coefficients ($C_{\eta,\xi_{E,..,S}}$, defined in B).  C: An example of cell-adhesion level for a keratinocyte at a grid point $\eta$. This keratinocyte happens to have one keratinocyte and one melanocyte neighbor,  from (\ref{Ad}) an adhesion value $A_\eta$ is determined ($A_\eta = 0.375$). D: This keratinocyte can move to either the north or south grid point (empty) with the same probability of $\mbox{Pr} = (1-A_\eta)/2 = 0.3125$. E: Another example, where a melanocyte happens to have three neighbors (east:melanocyte, south:keratinocyte, west:fibroblast) which gives a cell-adhesion value of 0.5. F: The probability of movement for this melanocyte is 0.5 (i.e. ($1-A_\eta$)) and the cell can move to either its north or south grid point. However, since there is an ECM gradient ($[E]$(south) $> [E](\eta)$ but $[E]$(north) $<$ $[E](\eta)$)the melanocyte is not allowed to move north and therefore can only move south. ECM gradient will always dictate the movement direction, provided one exists and their is sufficient space to move.}
\label{FigCAd}
\end{figure}

\vspace{.2cm}
\noindent{\em{Cell-cell adhesion}}: In order to incorporate one of the important mechanical aspects of skin into the model, we define an adhesion level ($ 0 \leq A_\eta \leq 1$) at each node $\eta \in \Omega$ explicitly by using cell-cell adhesive coefficients $C_{\eta,\xi}$, where $\xi$ is one of its orthogonal neighbors (i.e. $\xi \in \it{N}_\eta$). The adhesive coefficient between the same cell types is defined as 0.5. A value zero is assigned when there is no known adhesion mechanism between two cells, such as between a fibroblast and a keratinocyte. The coefficient for two melanocytes is based on current literature ~\cite{Haass:2005p80, Haass:2005p106}. To model E-cadherin mediated adhesion between a keratinocyte and a melanocyte, we impose a higher value (1) between these cell types.  To recapitulate the anchorage of a melanocyte to the basement membrane, we also assign a higher value (1). The adhesive coefficient $C_{\eta,\xi}$ for all cases is summarized in Figure~\ref{FigCAd} A-B.  By averaging the four adhesion coefficients $C_{\eta,\xi}$ at each grid point, we define an adhesion level $A_\eta$ for a cell at that grid point $\eta$. 
\begin{equation}
A_\eta =  \sum\limits_{\xi \in  N_\eta} C_{\eta, \xi} /4.0.
 \label{Ad}
\end{equation}

 \vspace{.2cm}
\noindent{\em{Migration}}: 
In vSkin, every cell has capacity to move to one of its orthogonal neighbors ($N_\eta$) regulated by its cell-cell adhesion restriction (i.e., $\mbox{Pr(migration) $\sim$ $(1-A_\eta)$}$), where $A_\eta$ is a cell-cell adhesion level obtained from equation (\ref{Ad}). Since the migration rules are slightly different for each cell type we will give a more detailed description in the following paragraphs.

Both keratinocytes and fibroblasts move only if empty grid points exist in $\it{N}_\eta$. We first determine the cell-cell adhesion value ($A_\eta$). If there is only one empty neighbor in the neighborhood $N_\eta$, the cell moves to this point with probability ($1-A_\eta$). If there are more empty grid points, the new position for the cell is chosen randomly from these grid points. An example of keratinocyte migration is shown in Figure~\ref{FigCAd}C-D.

In addition to the cell-cell adhesion restriction, we use more complex rules for melanocyte migration based on the current literature ~\cite{FITZPATRICK:1963fk,Haass:2005p106}. Melanocytes locate to the basement membrane in a 1:5 stable ratio with basal keratinocytes. Typically, one melanocyte is associated with approximately 36 keratinocyte neighbors. This symbolic structural relationship is known as the melanin unit. In the vSkin model, this melanin unit is modeled as a number of keratinocytes in the upper half of the Moore neighborhood of range $\gamma$ for each melanocyte. The value of $\gamma$ is set to 4 in vSkin. In vSkin, a melanocyte is moved to one of its orthogonal neighbors with regards to its cell-cell adhesion level, provided that (i) the melanin unit of the melanocyte is disrupted, (ii) there is an ECM gradient towards one of its neighbors, and (iii) at least one of its neighbors is either empty or only occupied by a keratinocyte. In order to satisfy (i), a check for the number of keratinocyte neighbors is made every cell time step. Rule (ii) is implemented because melanocytes are known to have a delicate cell-matrix interaction mechanism~\cite{NikolasKHaassKerianSMSmalleyLingL:2005p949,Morelli:1993fk} and migrate via haptotaxis. Based on rule (iii), a melanocyte is allowed to be moved to another grid point $\xi$ in $N_\eta$ even if the point is occupied by a keratinocyte. This additional rule is implemented to recapitulate observed melanocyte high motile activity~\cite{NikolasKHaassKerianSMSmalleyLingL:2005p949}. If there exist one neighbor ($\xi$) satisfying rule (ii) and (iii), the melanocyte is moved to the new position $\xi$. If more than one such neighbors exist, the new position for the melanocyte is chosen randomly from the neighbors. An example of melanocyte migration is provided in Figure~\ref{FigCAd}E-F.

 \vspace{.2cm}
\noindent{\em{Proliferation}}: In the vSkin model, we assume that each cell has capacity for proliferation and will produce two daughter cells provided that (i) the cell specific growth factor is sufficient for cell growth and (ii) there is sufficient space surrounding the parent cell for the two daughter cells to occupy. For a keratinocyte to divide, the concentration of $G_E$ has to be sufficient ($G_E > B_k$) for its division. Then, a keratinocyte proliferates with a probability of  $\mbox{Pr(keratinocyte division)} \sim (1-T_\beta)$. A melanocyte proliferates if (i) the melanin unit is disrupted and (ii) the concentration of bFGF is sufficient for cell division ($G_b > B_m$). A fibroblast is allowed to proliferate if the growth factor bFGF ($G_b$) is greater than the threshold ($B_f$). In addition to the bFGF level, we also consider a space constraint that limits the number of fibroblasts in vSkin domain. A fibroblast can divide only if at most one fibroblast neighbor exists in its neighborhood. This is not an unreasonable assumption in that the main role of fibroblasts in the model is to supply microenvironmental factors and the number of fibroblasts and the production rate of the microenvironmental factors are scalable. In other words, we can first choose a fixed number of normal fibroblasts in the model and then scale the production rates ($\alpha_E,~\alpha_b,~\alpha_{T_\beta}$) accordingly. In order to satisfy (ii), we assumed that one daughter cell replaces the parent cell and the other daughter cell will move to any one of the parent cell's four empty orthogonal neighbors ($N_\eta$). If more than one of the neighboring grid points is empty, then the new cell position is chosen randomly from these points. 

 \vspace{.2cm}
\noindent{\em{Death}}: For any cell to survive, it requires sufficient growth factors. We assume that the concentration has to drop to starvation levels, $S_k$ for a keratinocyte and $S_m$ for a melanocyte, before death can occur. For a normal fibroblast, we use a different constraint that incorporates both cell age and the concentration of bFGF. Since every fibroblast is continuously producing bFGF with the rate $\alpha_b$, it is never deprived of growth factor in normal skin conditions. We make another assumption that as a fibroblast ages, it will need more growth factors. In other words, the probability of fibroblast death is positively correlated with age but has a reciprocal relationship to the concentration of bFGF ($\mbox{Pr(fibroblast death)} \sim [G_b]/\mbox{age}$). The space that dead cells occupy immediately becomes available to new cells.

\subsubsection*{HCA Algorithm}

We now summarize the computational algorithm that integrates the discrete cells and the continuous microenvironmental variables.

\begin{enumerate}
\item Define a rectangular grid ($\Omega$) that determines both microenvironmental concentrations and cell positions.
\item Set initial concentration of EGF, bFGF, $\mbox{TGF}_\beta$ and ECM at each node $\eta \in \Omega$.
\item Place fibroblasts, melanocytes and keratinocytes on the grid.
\item Microenvironmental concentrations are solved using the governing equations (\ref{eq1}) - (\ref{eq5}).
\item Cell-cell adhesion level $A_\eta$ is determined for each $\eta \in \Omega$.
\item The concentration of microenvironmental factors and adhesion level are coupled with each cell to determine its phenotype.
\item Action associated with the determined phenotype is realized and the grid is updated accordingly.
\item Go to step 4. 
\end{enumerate}

\subsubsection*{Parameterization}
This vSkin model inevitably contains a large number of parameters as it has three different cell types and five microenvironmental factors. Unfortunately, many of the model parameters were difficult, if not, impossible to obtain, and therefore need to be estimated. However, due to the interdependent nature of the variables, precise parameterization may not be as important, rather the interactions between the variables is the real driving force. Homeostasis is our primary goal and as such served as central driver for parameterization. We do not doubt that other parameter sets could achieve similar outcomes. Note that all parameters are non-dimensionalized~\cite{Anderson:2005p10258}.

The diffusion rate of EGF ($D_{E}$) is estimated using the relation given in \cite{Swabb:1974p10166} with the molecular weight  133.07 kDa~\cite{molecularweight}. The estimated value is $D_{E} =  2.5 \times 10^{-8}\ \mbox{cm}^2/\mbox{s}$. The production rate of EGF by a fibroblast has not been measured. Thus, we estimate it to be a 0.5 (dimensional value = $1.2~\mbox{fg}~\mbox{day}^{-1}\mbox{cell}^{-1}$) from multiple simulations varying this parameter within range of [0,1] to achieve normal growth (net growth $\sim 0$) of keratinocytes. Fibroblasts are the main growth factor supplier in skin~\cite{Sorrell:2004bh,Kalluri:2006uq}. Thus, we make an assumption that a keratinocyte produces EGF with a significantly smaller rate (0.005; non-dimensional). Growth factor consumption rates are difficult to measure, especially at the level of single cells. We consider a normal activity rate of growth factor receptors. For example, we use 0.01 for a normal activation rate of EGF receptors of keratinocytes, meaning that a keratinocyte is able to bind 1\% of EGF at the grid point (per cell) per time step. The rate is set to be a smaller value (0.001) for melanocytes since EGF is known to stimulate keratinocyte growth mainly~\cite{Green:1985vn,Nanney:1986ys}. Note that we varied these consumption rates within a range $[0,1]$ to achieve skin homeostasis. We estimate the diffusion rate of $\mbox{TGF}_\beta$ using the relation in \cite{Swabb:1974p10166} and the molecular weight  44.3112 kDa~\cite{molecularweight}. The estimated value is $D_{T_\beta} = 5.8\times 10^{-8}\ \mbox{cm}^2/\mbox{s}$. The production rate of $\mbox{TGF}_\beta$ by a fibroblast is chosen to be 0.21333$~\mbox{fg}~\mbox{day}^{-1}\mbox{cell}^{-1}$ from {\it in vitro} study~\cite{Partridge:1989vn}. The rate for a keratinocyte is assumed to be significantly smaller than that of a fibroblast~\cite{Amjad:2007uq}. Since $\mbox{TGF}_\beta$ is known to mainly regulate the growth of keratinocytes~\cite{Krane:1991zr, Hannon:1994fk,Ewen:1996uq}, we assumed that keratinocytes bind to $\mbox{TGF}_\beta$ the most. The rates for a melanocyte and a fibroblast were estimated to achieve normal skin homeostasis. The consumption rates of $\mbox{TGF}_\beta$ are estimated to be  $ 0.2, \ 0.6, \ 0.1$ for a melanocyte, a keratinocyte, and a fibroblast, respectively. Using the same method~\cite{Swabb:1974p10166} with molecular weight 17.1531 kDa, the diffusion rate $D_b$ is estimated to be $1.2 \times 10^{-7} \ \mbox{cm}^2/\mbox{s}$. The production rate of bFGF by a fibroblast is also estimated from multiple simulations varying these parameters within a range of [0,1] (non-dimensional). The estimated parameter is 0.05 (0.12 $\mbox{fg}~\mbox{day}^{-1}\mbox{cell}^{-1}$). The rate for keratinocytes is assumed to be significantly smaller, since keratinocytes appear to produce significantly less bFGF than fibroblasts ~\cite{{Halaban:1988qf}}. Since both melanocytes and fibroblasts use bFGF as a growth promoting signal and there is no evidence showing which cell binds to bFGF more, we use the same value (0.02) for the bFGF binding rate. The rate for keratinocytes is set to be significantly smaller (0.005). The diffusion rate of MMP is estimated to be $D_p= 5.01 \times 10^{-8} \ \mbox{cm}^2/\mbox{s}$ in \cite{Swabb:1974p10166, molecularweight}. The production rate of MMP by fibroblasts is varied within a range 0.024 - 0.12 $~\mbox{fg}~\mbox{day}^{-1}\mbox{cell}^{-1}$. Estimates for the kinetic parameters $\lambda$ and $\mu$ are not available since they are very difficult to obtain experimentally. Thus, we set non-dimensional values that best produced homeostasis. We know that ECM density in the dermis is significantly higher than that in the epidermis (Figure \ref{Fig1}), we therefore set the epidermal ECM density to 0.2 and the dermal ECM density is 0.7. All of the parameter values described here are summarized in Table~\ref{tab2}.

\begin{table}[!h]
\caption{
\bf{Model parameters}}
\begin{tabular}{|cccc|}\hline
Parameter & Description &  Value & Normalized value \\ \hline
 $D_{T_\beta}$ & Diffusion rate of $\mbox{TGF}_\beta$  & $5.8\times 10^{-8} \ \mbox{cm}^2/\mbox{s}$\cite{Swabb:1974p10166, molecularweight}  & 0.0838\\
 $\alpha_{T_\beta}$ & Fibroblast $\mbox{TGF}_\beta$ production rate & 0.21333 $\mbox{fg }\mbox{day}^{-1} \mbox{cell}^{-1}$ ~\cite{Partridge:1989vn}& 0.0873 \\
 $\beta_{T_\beta}$ &  Keratinocyte $\mbox{TGF}_\beta$  production rate & $0.0106 \mbox{ fg }\mbox{day}\mbox{ cell}^{-1}$  & 0.004 \\    
$\gamma_{T_\beta}$ &  Keratinocyte $\mbox{TGF}_\beta$  binding rate & $ - $ & 0.6 \\
  $\zeta_{T_\beta}$ &  Melanocyte $\mbox{TGF}_\beta$  binding rate & $ - $  & 0.2\\
 $\xi_{T_\beta}$ &  Fibroblast $\mbox{TGF}_\beta$  binding rate & $ - $ & 0.1 \\
$D_{E}$ & Diffusion rate of EGF  & $2.5 \times 10^{-8} \ \mbox{cm}^2/\mbox{s}$\cite{Swabb:1974p10166, molecularweight} & 0.0367\\
 $\alpha_{E}$ & Fibroblast EGF production rate & $1.2 \mbox{ fg }\mbox{day}\mbox{ cell}^{-1}$  & 0.5\\
 $\beta_{E}$ &  Keratinocyte EGF production rate & $0.0012\mbox{ fg }\mbox{day}\mbox{ cell}^{-1}$  & 0.005\\
  $\gamma_{E}$ &  Keratinocyte EGF  binding rate & $ - $  & 0.01 \\
  $\zeta_{E}$ &  Melanocyte EGF binding rate & $ - $   & 0.001\\ 
 $D_{b}$ & Diffusion rate of bFGF  & $1.2 \times 10^{-7} \ \mbox{cm}^2/\mbox{s}$\cite{Swabb:1974p10166, molecularweight} & 0.1708\\
 $\alpha_{b}$ & Fibroblast bFGF production rate & $0.12 \mbox{ fg }\mbox{day}\mbox{ cell}^{-1}$  & 0.05 \\
 $\beta_{b}$ &  Keratinocyte bFGF production rate & $0.024\mbox{ fg }\mbox{day}\mbox{ cell}^{-1}$ & 0.01 \\
  $\xi_{b}$ &  Fibroblast bFGF  binding rate & $ - $ & 0.02 \\
  $\gamma_{b}$ & Keratinocyte bFGF binding rate & $ - $  & 0.005\\ 
  $\zeta_{b}$ &  Melanocyte bFGF binding rate & $ - $  & 0.02\\ 
   $\lambda$ & Decay rate of all growth factors &  - & 0.01\\
$D_{p}$ & Diffusion rate of MMP & $5.01 \times 10^{-8} \ \mbox{cm}^2/\mbox{s}$\cite{Swabb:1974p10166, molecularweight}  & 0.0724\\
 $\kappa_E$& Keratinocyte ECM production rate & $ 15.7 \mbox{ pg }\mbox{day}^{-1}\mbox{ cell}^{-1} $ & $0.1$\\
  $\rho_E$& Fibroblast ECM production rate & $ 15.7 \mbox{ pg }\mbox{day}^{-1}\mbox{ cell}^{-1} $ & $0.1$\\
  $\mu$ &  ECM decay rate & - & $0.001$\\ 
 $E_0$ & Epidermis ECM density & - & $0.2$\\
 $E_1$ & Dermis ECM density  & - & $0.7$ \\
 \hline
\end{tabular}
\label{tab2}
 \end{table}

The cell cycle time is set it to be a typical value $t_c = 24~h$.  The size of the grid is set to be $300\times 62$. The grid size $\delta h$ is set to be $\delta   h = 20~ \mu\mbox{m}$ (a typical cell diameter), and thus the dimensions of the skin slice we simulate are $6 ~\mbox{mm}  \times 1.24~ \mbox{mm} $. The time step for the microenvironmental variables is set to be $\delta t = 4.8 ~h$.  We assume that all cell types have the same volume $3.1416\times10^{-9}~\mbox{cm}^3$ with a radius $10 ~\mu\mbox{m}$, and therefore the base number of cells is set to be $3.1831\times 10^{8}$. The background density of ECM is set to be $5.0 \times 10^{-4} \mbox{g }\mbox{cm}^3$~\cite{Kim:2010fk}. The background concentration of microenvironmental variables is assumed to be $G_0 = 10 \mbox{ ng }\mbox{cm}^{-3}$~\cite{Tavakkol:1999fk}. We assume that the concentration of growth factors above which a cell proliferates in normal skin is set to the non-dimensional threshold value 0.02 for $B_m$, 0.02 for $B_f$ and 0.01 for $B_k$. Since the concentrations below which cells die from starvation are also variables, we set a starvation threshold to be 1\% ($S_k$), 2\% ($S_m$), and 0.1\% of the background microenvironmental concentration $G_0$ for keratinocyte, melanocyte, and abnormal fibroblast respectively. 
  
\subsection*{Experimental methods}

\subsubsection*{Cell culture}
 Melanoma cells were kindly provided by Dr Meenhard Herlyn (The Wistar Institute, Philadelphia, PA) and were cultured in RPMI medium supplemented with 5\% fetal calf serum (FCS).

\subsubsection*{Co-culture growth}
 Senescence was induced in primary fibroblasts following irradiation with 10 Gy followed by 7 days of recovery as described in Campisi and colleagues~\cite{Coppe:2008p7159}. Senescent and pre-senescent fibroblasts were plated upon 10 cm tissue culture plates overnight before the addition of Adeno-GFP tagged melanoma cells (a gift from Dr Amer Beg, Moffitt Cancer Center). Images of the growing cells were taken after 24, 48 and 72 hours with a Nikon-TS100 inverted fluorescence microscope. 

\subsubsection*{Western blotting}
Cells were plated and grown overnight. To detect ADAM-9 levels, ADAM-9 antibody (ab) was used for immunoblotting (IB). For all IB experiments, proteins were denatured prior to separation on 6-18\% Tris-Glycine gels. Proteins were transferred to polyvinylidene fluoride (PVDF) and blocked with 5\% non-fat milk in Tris-Buffered Saline Tween-20 (TBST). Blots were incubated overnight in primary ab diluted according to the manufacturerÕs instructions.  Secondary ab's conjugated to the enzyme horseradish peroxidase (HRP) were used for detection by chemiluminescense.

\subsubsection*{Zymography}
Cells were plated and allowed to grow overnight. To detect functional ADAM-9 activity, proteins were not denatured prior to separation upon a casein gel. Cleavage of casein was imaged using an Epson V300 photo scanner.  

\subsubsection*{Three-dimensional spheroid assays}
WM793 melanoma spheroids were prepared using the liquid overlay method~\cite{Smalley:2006fk}. Spheroids were treated with serum free RPMI, pre-senescent fibroblast conditioned RPMI or senescent fibroblast conditioned RPMI. Pictures of the invading spheroids were taken using a Nikon-TS100 inverted fluorescence microscope. Images were converted to gray scale and the spheroid frames were outlined using ImageJ (Image Processing and Analysis in Java) by modifying the threshold. The spheroid frames were than inverted and absolute intensity of the invading cells was quantified using Adobe Photoshop.   

\subsubsection*{Human specimen procurement}

A representative tumor sample was obtained from a patient who underwent surgical resection for metastatic melanoma and was prospectively consented and accrued to an existing melanoma tissue procurement protocol approved by the Moffitt Cancer Center Scientific Review Committee and The University of South Florida Institutional Review Board. After deparaffinization, the slides were stained for ADAM9 (Chemicon) before a thorough washing and incubation with a Alexa-Fluor 647 secondary antibody. The slide was also stained with DAPI to indicate the nuclei and analyzed using a Leica confocal microscope. Areas of stroma, tumor and leading edge were delineated through examination of corresponding H \& E stained slide by a dermatopathologist.

\begin{figure}[!ht]
\begin{center}
\includegraphics[width=4in]{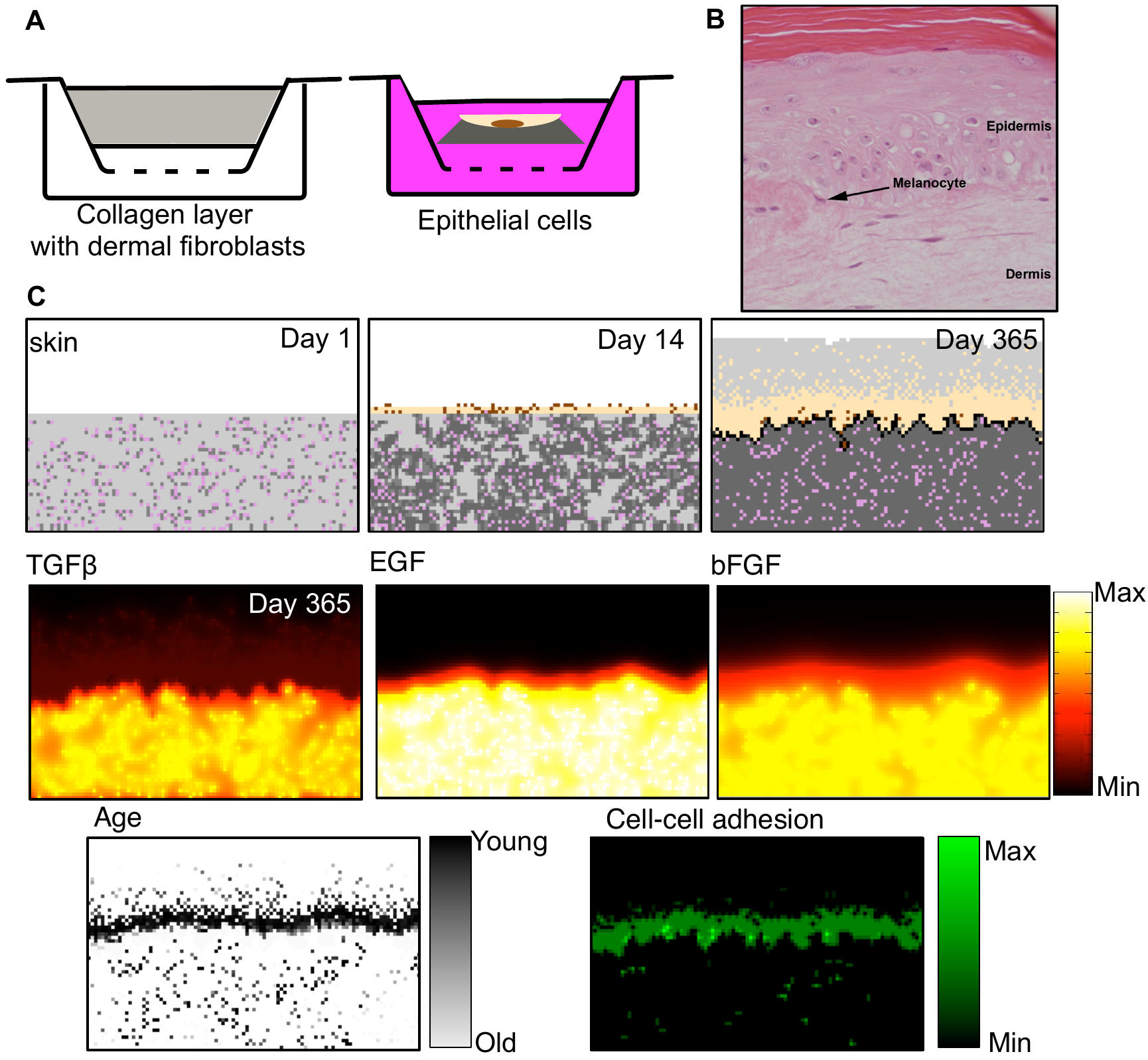}
\end{center}
\caption{
{\bf Organotypic skin and vSkin development.} A: Scheme showing the generation of {\it in vitro} orgonotypic skin reconstructs. A stromal layer of collagen mixed with fibroblasts is placed on top of the acellular collagen, and the layer is incubated for 4 \-- 7 days to allow the fibroblasts to constrict the collagen. The epithelial cells are layered on top of the fibroblast-containing stromal layer, and then the reconstruct is air lifted to promote differentiation and stratification of the epithelial layer. B: Photomicrograph shows a mature organotypic culture of a normal skin. C: Snapshots of vSkin initialization. The steps of the simulation follow the same experimental steps used to develop organotypic skin reconstruct. vSkin development at time step 0, 14 and 365 days are shown in the first row, and the distributions of growth factors, age and force at the end point (t = 365 days) are shown in second and third rows. }
\label{Fig4}
\end{figure}

\subsubsection*{Initialization}

Using the HCA algorithm and parameterization described above, we first ran an initial simulation to obtain the starting configuration of the domain (Figure~\ref{Fig1} B) for all subsequent simulations. We follow the experimental steps in the {\it in vitro} 3D organotypic skin reconstruct experiment Figure \ref{Fig4} A-B, as if we develop a virtual organotypic skin reconstruct. This 3D organotypic culture system has served as a foundation for many basic science studies as well as a skin transplantation model, and it is known to be very stable and homeostatic (see review in~\cite{Brohem:2011fk}). The initial simulation starts with a dermal layer mixed with fibroblasts that we simulate for two weeks until dermal fibroblasts produce enough growth factors and ECM for keratinocyte and melanocyte growth. Then a mixed population of keratinocytes and melanocytes are placed on the top of this matured dermis. As soon as keratinocytes are placed, they rapidly grow until they fill almost the entire domain. Soon after the most of keratinocytes become quiescent and start to die. Finally, the population becomes stabilized. The melanocyte population also rapidly increases initially but soon finds its equilibrium. This initial simulation shows the key interactions in the model are well balanced and manifest a normal skin homeostasis.  Figure~\ref{Fig4}C shows how the skin structure develops over the period of a year as well as the concentrations of the microenvironemental factors, the age of all cells and the adhesion field after one year. After a period of around 6-8 weeks, vSkin has already reached a homeostatic configuration. Reassuringly, the timescales of vSkin formation and the organotypic skin reconstruct are similar. We will utilize the resulting skin structure that emerges after a year as our initial condition in all subsequent simulations, since the skin structure that naturally emerged from this initial simulation is morphologically close to real skin.

\subsubsection*{Skin fitness}
The degree of abnormality of vSkin was characterized by a defined metric, ``skin fitness ($f(t)$)."  We quantify ratios of each cell compartment (ratio of keratinocyte to melanocyte and that of fibroblasts to melanocytes) at each time step ($t$) and compare with the ratios of the normal state ($t_0$). We also took changes of epidermal thickness into account to monitor skin fitness. Changes from time step $t_0$ to time step $t$ ($t_0\rightarrow t$) in both compartmental ratios and epidermal thickness were weighted equally to determine the final skin fitness. The skin fitness is scored from -1 to +1, where +1 represents the maximal fitness and a normal skin state, and is defined as follows, 
\begin{equation}
{f(t)} = 1.0 - \left( w_1\frac{\Delta r}{r} + w_2\frac{\Delta h}{h}\right),
\label{skinFit}
\end{equation}
where $r$ represents the sum of two ratios, the ratio of keratinocyte to melanocyte and that of fibroblast to melanocyte, $\Delta$ represents the change from time step $t_0$ to $t$, $h$ stands for a epidermal thickness, and  $w_1$ and $w_2$ are the weights. In this study we choose the same weight (i.e.,$w_1 = w_2 = 0.5$).

\begin{figure}[!ht]
\begin{center}
\includegraphics[width = 4in]{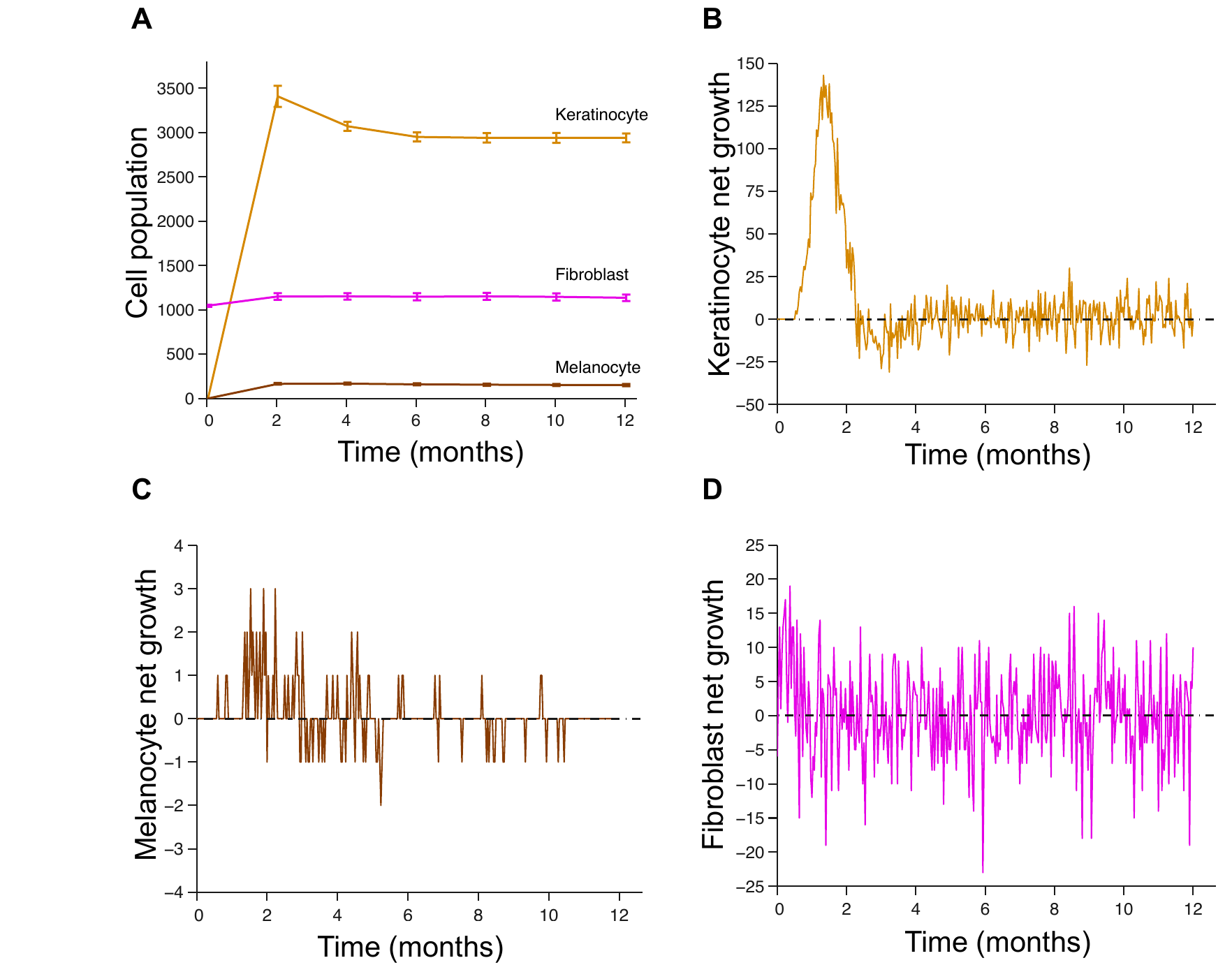}
\end{center}
\caption{
{\bf Temporal evolution of each cell population.} A: Mean populations dynamics of keratinocytes, melanocytes and fibroblasts (out of 50 realizations). Both keratinocyte and melanocyte populations initially increase after placement in the vSkin domain, but after two months the population is stabilizes. B: A sample of keratinocyte net growth over one year. Keratinocytes proliferate until the population reaches a maximum in the vSkin domain, then after about two months, keratinocytes start to commit cell death and the net growth begins to oscillate around zero. C: A sample of melanocyte net growth over one year. Initially, there are more births than deaths, but soon after the net growth oscillates around zero. D: A sample of fibroblast net growth over one year.}
\label{Fig5}
\end{figure}

\section*{Results}

\subsection*{The vSkin model maintains a stable homeostasis}
We already know that our vSkin model can recapitulate normal skin structure. As a typical CA method inevitably contains stochastic components, multiple realizations ($ \geq 50$) were carried out for each simulation.  Our simulations show that vSkin can reach a stable cell net growth rate and total cell number in the domain.  Keratinocytes rapidly grow for the first two months and then reach a stable state (Figure~\ref{Fig5} A and B), while both melanocytes and fibroblasts retain their initial population and the net growth remains stable (Figure~\ref{Fig5} A,C, and D). These data indicate that our vSkin model rapidly achieves and maintains a stable skin homeostasis.
\begin{figure}[!ht]
\begin{center}
\includegraphics[width = 4in]{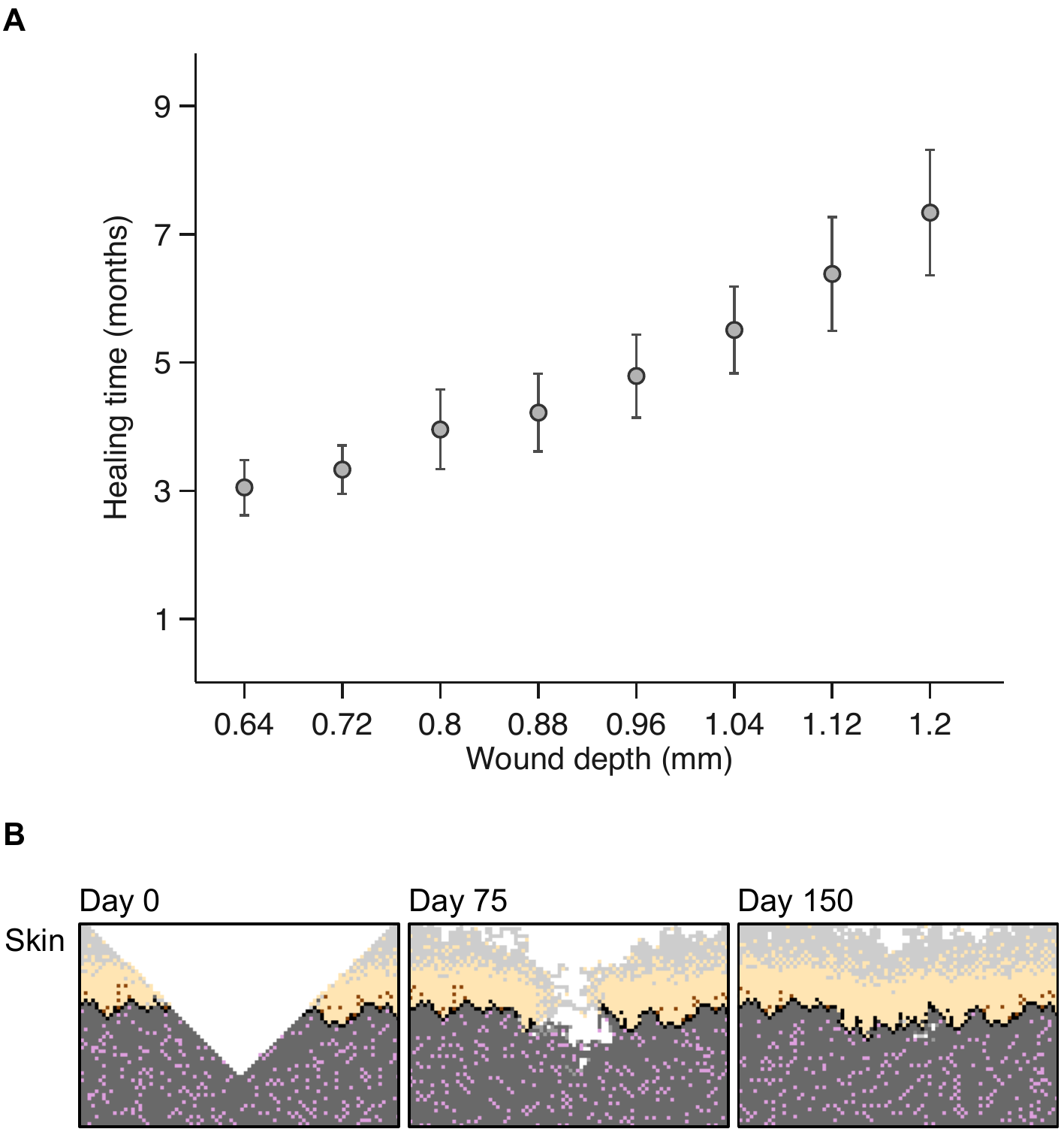}
\end{center}
\caption{
{\bf Wound healing.} A: Positive correlation between healing time and wound size (depth) derived from a suite of wound healing simulations with a range to different wound depths (average of 100 runs). The healing time was defined as the time it takes for the basement membrane to be continuously connected. The mean healing time increases as wound size increases. B: Three snapshots from a wound healing simulation, with a wound that was generated by removing a triangular shape slice with height of 0.96 mm and width of 1.86 mm.}
\label{Fig6}
\end{figure}

\subsection*{The vSkin model is robust to physical perturbation}
Once the stability of the vSkin model has been established, we next examined its robustness. Two different types of perturbations were applied to determine the robustness of our model system. First, we tested if our vSkin model can withstand massive loss of its constituents. To this end, a triangular shaped injury, which mimics an {\it in vivo} puncture wound to the skin, was created in the center of domain. As a new space was introduced into vSkin, cells nearby the space have an increased tendency to move toward the new space  due to gradient in the force distribution. In the dermal layer, fibroblasts migrate into the injury site and start to produce growth factors (EGF, $\mbox{TGF}_\beta$, and bFGF) and extracellular matrix proteins. These newly produced growth factors promote growth of keratinocytes in epidermis, thereby starting to fill the void (healing procedure). A new basement membrane is generated by interactions between fibroblasts and either keratinocytes or melanocytes. The recovery time was defined as the time it took the wound site to form a complete basement membrane. 
 
 We also changed the size of the wound by increasing injury depth, and quantified the relationship between healing time and injury size. The recovery time from wounding tends to increase with wound size (see Figure~\ref{Fig6} A). One hundred realizations were carried out and the mean and standard errors are reported. Interestingly, the variation among realizations increases with wound size. This is presumably due to the stochastic nature of the cell migration, thereby in basement membrane generation. In other words, the  development of new basement membrane is dependent on chance interactions between fibroblasts and epidermal cells (keratinocytes or melanocytes).  An example snapshot of the wound-healing process is depicted in Figure~\ref{Fig6} B. These wound healing simulations indicate that our vSkin model is robust in the face of significant physical perturbation. Another implication is that the vSkin model can recapitulate some basic {\it in vivo} skin dynamics. 
\begin{figure}[!ht]
\begin{center}
\includegraphics[width = 5in]{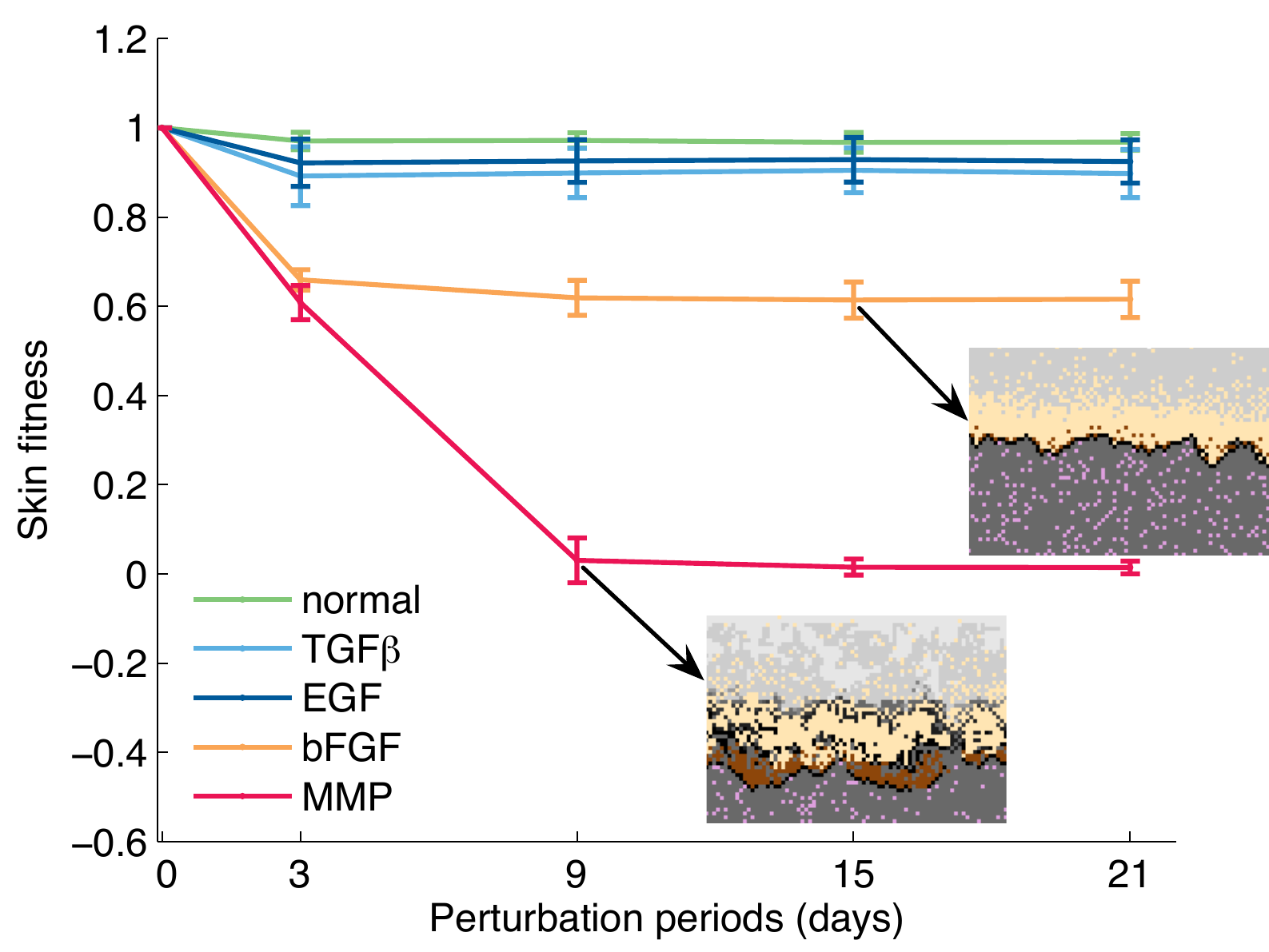}
\end{center}
\caption{
{\bf Growth factor perturbations and skin fitness.} Simulation results from maximizing (1.0) the concentration of each growth factor over the entire domain during 3, 9, 15 and 21 days (averaged over 50 runs). After the perturbed skin reaches its new equilibrium, the skin fitness was measured using (\ref{skinFit}). The EGF and $\mbox{TGF}_\beta$ impulses have a negligible effect on skin fitness. The bFGF and MMP have a greater impact on skin fitness, reducing it to 60\% and 0\%, respectively. However, whilst the skin structure is quite close to normal (see upper inset vSkin histology) after bFGF perturbation, after MMP over-expression, the skin has quite a different structure (see lower inset vSkin histology)characterized by a lower basement membrane, and the epidermal cells (keratinocytes and melanocytes) have populated the dermal layer. }
\label{Fig7}
\end{figure}

 \subsection*{The vSkin model returns to equilibrium after biochemical perturbations}
The second perturbation involves manipulation of  microenvironmental factors to determine if the vSkin system is robust enough to withstand super-physiological microenvironmental changes. Specifically, each growth factor was maximized in the domain during 3, 9, 15 and 21 days. The skin fitness function $f(t)$ in equation (\ref{skinFit}) was used to quantify abnormality of vSkin after perturbation. Note that we measure skin fitness $f(t)$ when vSkin reached a new equilibrium.
 
When EGF impulses were given to vSkin, keratinocytes responded very quickly. Keratinocytes proliferated until their population saturated the domain and then became quiescent. After most of EGF has either decayed or been consumed by keratinocytes and melanocytes, the quiescent keratinocytes deprived of EGF start to die. The resulting skin histology appears to be normal as it preserves normal epidermal thickness and the normal ratio of melanocytes to keratinocytes. In Figure~\ref{Fig7}, the blue line shows what little impact the EGF impulse made on vSkin, even for the longest perturbation time.

The effects of $\mbox{TGF}_\beta$ impulses on vSkin were also negligible even though the growth of keratinocytes was inhibited by high $\mbox{TGF}_\beta$ concentration. This is presumably due to fast turn over rate of keratinocytes. The light blue line in Figure~\ref{Fig7}  depicts how the skin fitness changes over different periods of  $\mbox{TGF}_\beta$ perturbation.

Upon applying bFGF impulses, melanocytes proliferated faster and started to occupy more space than keratinocytes. The overgrowth of melanocytes resulted in inhibition of growth of keratinocytes due to a lack of space. When overgrown melanocytes have consumed the excess bFGF, they started to commit apoptosis. Unlike EGF or $\mbox{TGF}_\beta$ impulses, vSkin obtained a new equilibrium state with a slight increase of melanocyte number, particularly near the basement membrane. The overall fitness score is 0.6 as depicted by the orange line in Figure~\ref{Fig7}.

\subsection*{Microenvironmental factors can transform normal skin to pathologic skin }
Unlike EGF, $\mbox{TGF}_\beta$, or bFGF, the overexpression of MMP had  profound impact on vSkin fitness. Upon applying MMP, the extracellular matrix and the basement membrane begin to degrade. Destruction of the basement membrane increased the probability of melanocytes and keratinocytes to migrate into the dermal layer. In particular, melanocytes had a higher tendency to migrate into the dermal layer (a new microenvironment for them) where they can find more growth factors and loss of control from keratinocytes. This new microenvironment stimulates rapid growth of melanocytes. As a result of MMP overexpression, the vSkin became thinner with both keratinocytes and melanocytes moving into the dermal layer (downward). vSkin was able to recover from small changes induced by short term MMP overexpression (3 days), but fails to compensate for long-term MMP overexpression (9, 15, and 21 days), as shown in Figure~\ref{Fig7}.
 
Collectively, the results indicate our vSkin model is able to return to its original homeostatic state or to find a new equilibrium in the face of significant microenvironmental deviations; however, this is dependent upon the perturbations only being to those variables present in normal skin (e.g., EGF, $\mbox{TGF}_\beta$, and bFGF). When vSkin was disturbed by a factor not typically present in normal skin, its robustness is dependent on the duration of exposure. With long term exposure of MMP, vSkin transforms to a pathologic state.  This implies that the regulation of microenvironmental factors contributes to vSkin transformation from a normal to an abnormal, pathologic state. 
\begin{figure}[!ht]
\begin{center}
\includegraphics[width = 4in]{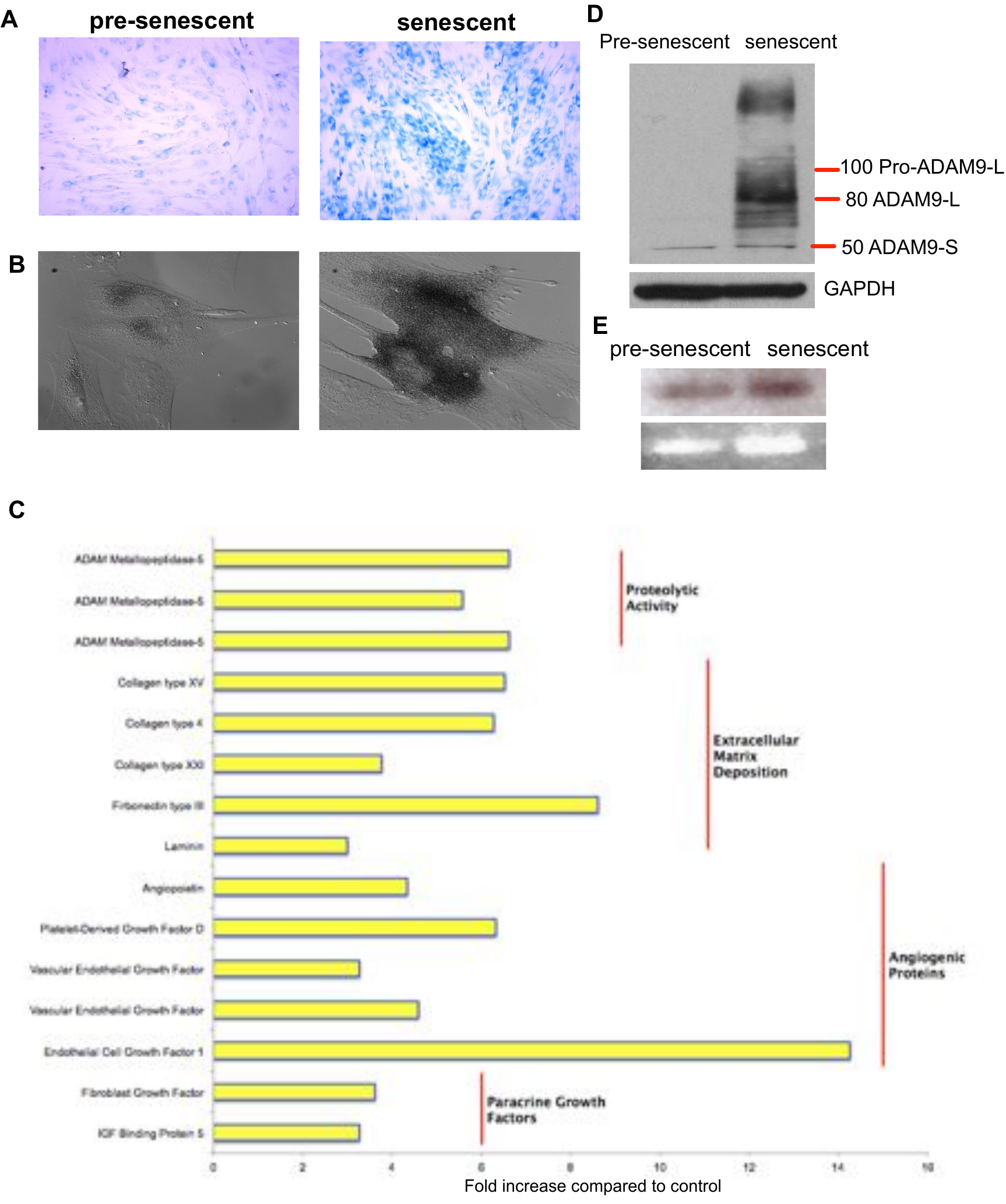}
\end{center}
\caption{
{\bf Phenotype changes in fibroblasts.} A: SA-$\beta$-Galactosidase staining where the blue cells are SA-$\beta$-Galactosidase positive (indicating senescence). B: Automated inverted fluorescence images of pre-senescent (left) and senescent (right) fibroblasts. C: Micro-array analysis shows the fold increase in expression of multiple proteins in senescent fibroblasts compared to normal fibroblasts. Specific families of growth factors, proteases and extracellular matrix protein show increased expression. D: Senescent fibroblasts showed enhanced expression of ADAM9 proteins. E: Zymography shows senescent fibroblasts enhance matrix degrading capacity. }
\label{ExpFig1}
\end{figure}

\begin{figure}[!h]
\begin{center}
\includegraphics[width = 3in]{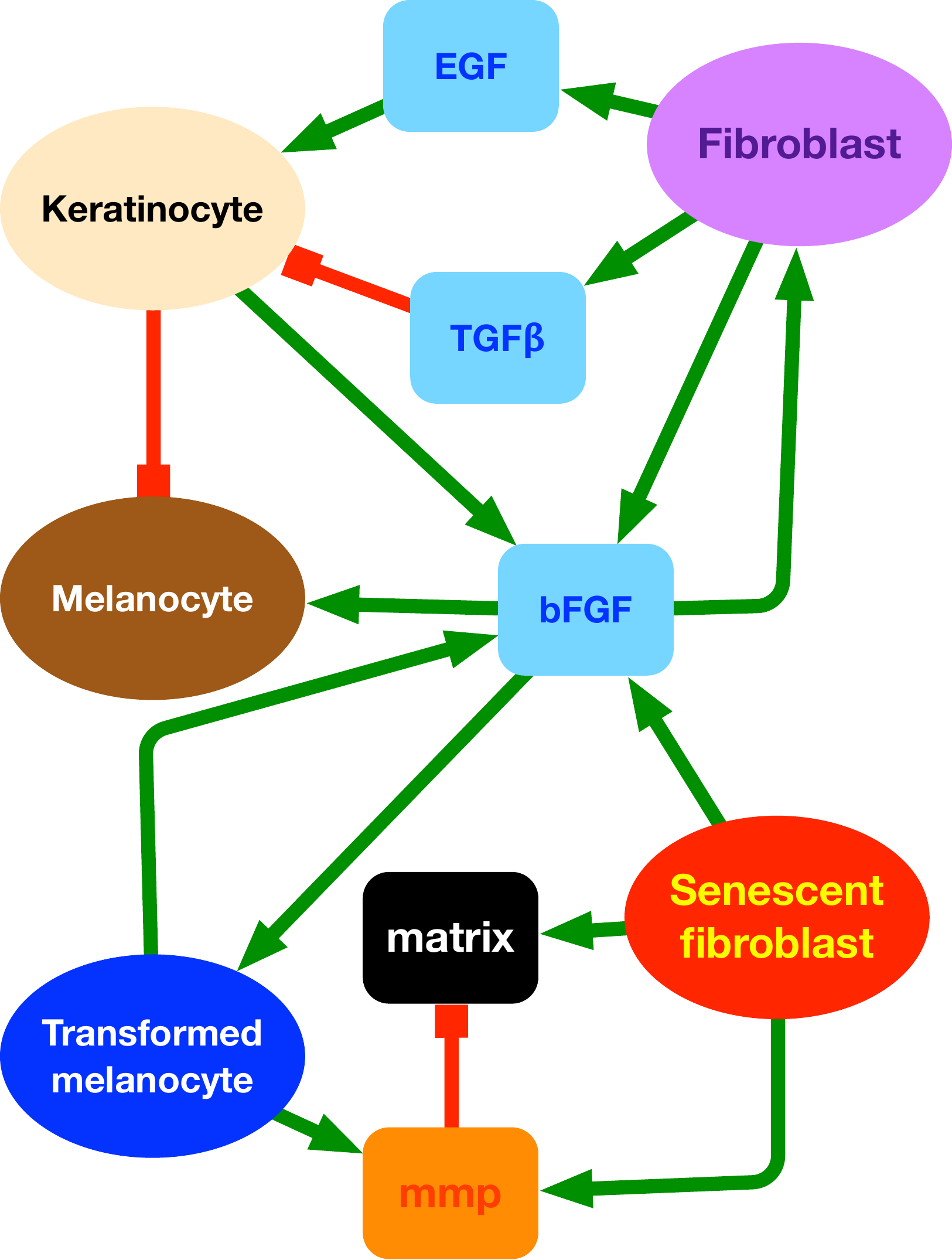}
\end{center}
\caption{
{\bf Cell interaction network for abnormal skin.} Key interactions among normal cells (keratinocyte, melanocyte, fibroblasts), abnormal cells (senescent fibroblast and transformed melanocyte) and microenvironmental variables represented as network. Note, cell coloration is used in all simulations to represent the equivalent cell. Green lines represent promoting events, while red lines imply inhibiting effects. A transformed melanocyte can produce its own growth factor (bFGF) and MMP, while senescent fibroblasts can produce bFGF, MMP and extracellular matrix proteins.}
\label{Fig8}
\end{figure}

\subsection*{Homeostatic disruption can lead to melanoma initiation}
Our {\it in silico}  perturbations have shown that vSkin is a robust system which recapitulates normal skin form and function. These experiments have also shown that bFGF and MMP are important microenvironmental regulators in skin homeostasis, particularly in melanocyte homeostasis. 
Melanocytes are the cell of origin for one of the most malignant forms of skin cancer, melanoma. A key to understanding melanoma initiation is determining how transformed melanocytes disrupt skin homeostasis. As we have readily shown in the previous section this homeostasis is crucially dependent on both keratinocyte regulation of melanocytes as well as microenvironmental regulation. Fibroblasts are the primary source for many of the microenvironmental factors that regulate skin homeostasis. Therefore, disruption or transformation of fibroblasts should also have a significant impact on skin homeostasis. Intriguingly, the major characteristic of transformed melanocytes that distinguishes them from normal melanocytes is their ability to produce bFGF and MMP.  Similarly, we now know that senescent fibroblasts show altered patterns of ECM expression, growth factor release and protease activity~\cite{Coppe:2008p7159}. 

To investigate this secretory phenotype further we undertook our own experimental investigation of the changes that occur in fibroblasts when they become senescent. Irradiation (10 Gy) of human primary skin fibroblasts led to their entry into a senescent-like state, as demonstrated by increased staining for $\beta$-galactosidase (Figure~\ref{ExpFig1}A,B). The onset of senescence in the fibroblasts was associated with a marked change in their mRNA profiles with increases noted in ECM constituents, growth factors and proteases (Figure~\ref{ExpFig1}C). Of these, ADAM9 is a protease thought to be critical for melanoma/fibroblast interactions~\cite{Zigrino:2011ly}. Western blotting demonstrated that ADAM9 expression was highly upregulated in senescent fibroblasts compared to non-senescent fibroblasts and that this was associated with increased matrix degrading capacity (Figure~\ref{ExpFig1}D,E). Therefore fibroblasts produce protease, increased levels of bFGF, and extracellular matrix proteins as they become senescent. 

In order to explore the impact of modifying these cell types on homeostasis, we incorporated transformed melanocytes and senescent fibroblasts into the vSkin model. Figure~\ref{Fig8} shows how these altered cell phenotypes modify the cell interaction network, now incorporating normal skin cells, transformed melanocytes and senescent fibroblasts. 

\begin{figure}[!ht]
\begin{center}
\includegraphics[width = 3in]{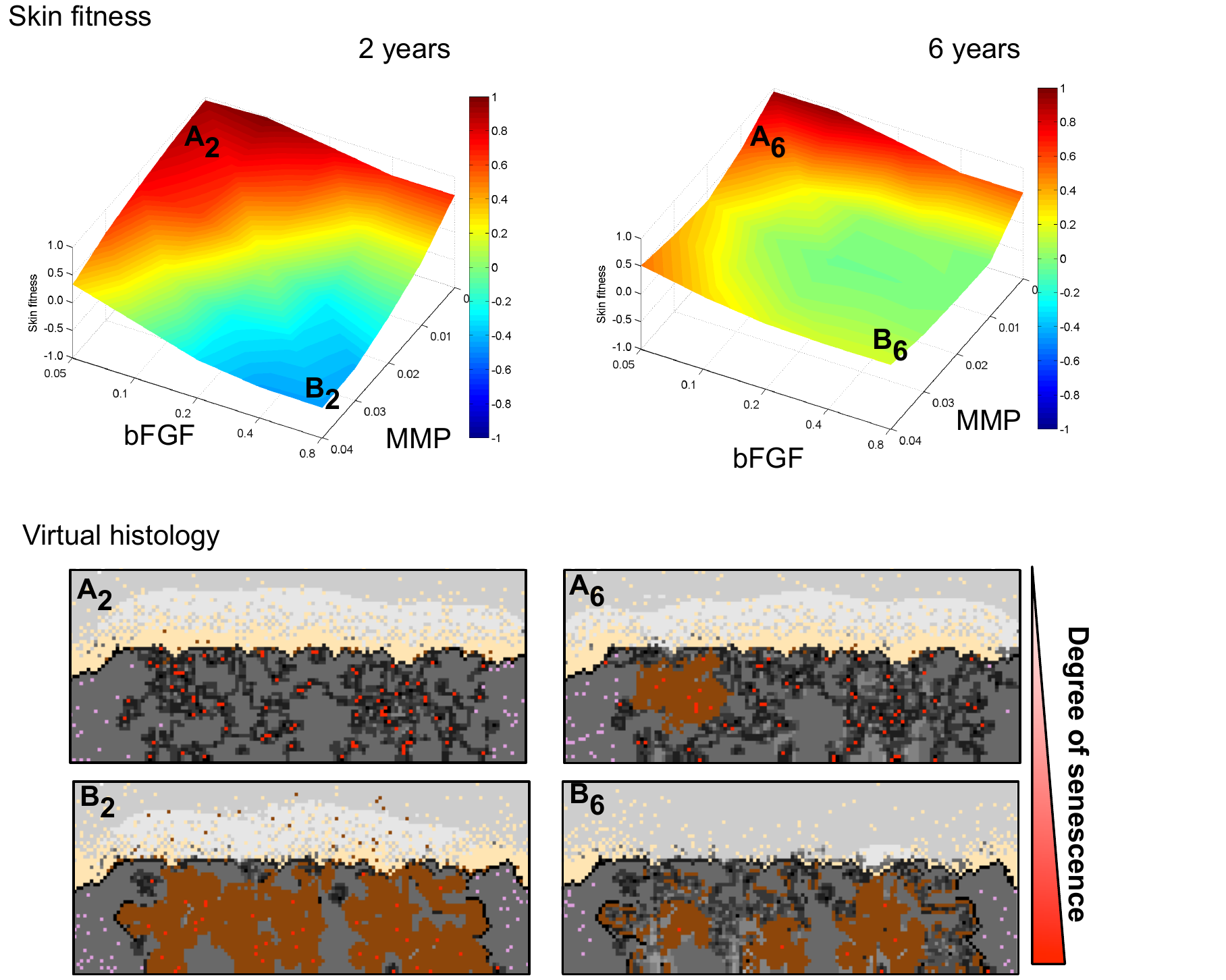}
\end{center}
\caption{
{\bf Degree of senescence and its effects on melanocyte growth.} Senescent fibroblasts were placed in the center of the dermal layer, among normal fibroblasts, only 10\% were chosen to be senescent. We considered a range of senescence in terms of the production rate of bFGF and that of MMP. Specifically, the nondimensional parameter space for senescent fibroblasts was $(\alpha_{b}^s,\alpha_{p}^s) \in \{ (x,y)|  0.05 \leq x \leq 0.8,  \   0.0 \leq y \leq 0.04 \}$, where $\alpha_{b}^s$ represents bFGF production rate by senescent fibroblasts and $\alpha_{p}^s$ indicates MMP production rate by senescent fibroblasts. Multiple realizations (50) with each parameter set  $(\alpha_{b}^s, \alpha_{p}^s )$ were performed, and the mean skin fitness was quantified at each time step using (\ref{skinFit}). The first row shows the skin fitness landscapes after two and six years. A2 and A6 represent skin fitness obtained from simulations with mildly senescent fibroblasts (bFGF production rate is 0.05 and MMP production rate is 0.01). B2 and B6 show the fitness with highly senescent fibroblasts (bFGF production rate is 0.8 and MMP production rate is 0.040). Lower panels show representative vSkin histology for each of these points and highlights how different they are from normal skin (see figure \ref{Fig8} for a key of color coded cell types).}
\label{Fig9}
\end{figure}

\subsubsection*{Senescent fibroblasts aid melanocyte proliferation and invasion}
To tease apart the relative contributions of the two abnormal cell types we initially look at each of them separately. First, we studied the role of senescent fibroblasts by examining how the degree of senescence in a subpopulation of the fibroblasts transforms normal skin structure (Figure~\ref{Fig9}). The degree of senescence specifically refers to how much bFGF and MMP the fibroblasts produce, so the most senescent phenotype produces maximal bFGF and MMP. For each parameter set (bFGF production rate, MMP production rate), we carried out multiple realizations and quantified skin fitness  at each time step ($f(t)$) using the average number of each cell compartment and the average epidermal thickness from these realizations. We observed that any degree of senescence for the fibroblasts significantly enhanced growth of melanocytes and promoted invasion into the dermis. From the skin fitness landscapes (upper panel, Figure~\ref{Fig9}) it is clear that the intensity of skin disruption is directly dependent on the degree of fibroblast senescence.  When senescent fibroblasts have weak secretory phenotypes (i.e., close to normal), the fitness level of the skin they produce is almost one (i.e., normal). The skin structure, however, is not completely normal as there is some accumulation of matrix near the senescent fibroblasts (see virtual histology Figure~\ref{Fig9}, $A_2$). When senescent fibroblasts have stronger secretory phenotypes ($B_2$ and $B_6$) the melanocyte population expands rapidly and skin fitness deteriorates to -0.4 ($B_2$). Interestingly,  the fitness subsequently increases to 0.0 as time progresses (see the difference between $B_2$ and $B_6$). This is due to the rapid growth of melanocytes initially $(B_2)$, driven by the senescent fibroblasts, but this excessive growth leads to deprivation of the growth factors for both fibroblasts and keratinocyes (see Figure~\ref{Fig8} for the interdependence of cell types and growth factors) resulting in cell death. Whilst not cancer, excessive melanocyte production is a key feature of melanoma growth. It is important to note that the melanocytes here are normal and are only responding to the abnormal microenvironmental cues being produced by the senescent fibroblasts. However, if these melanocytes were to become transformed then this would be true melanoma, therefore, senescent fibroblasts may play a critical role in creating an environment ripe for melanoma initiation.
 
\begin{figure}[!ht]
\begin{center}
\includegraphics[width = 3in]{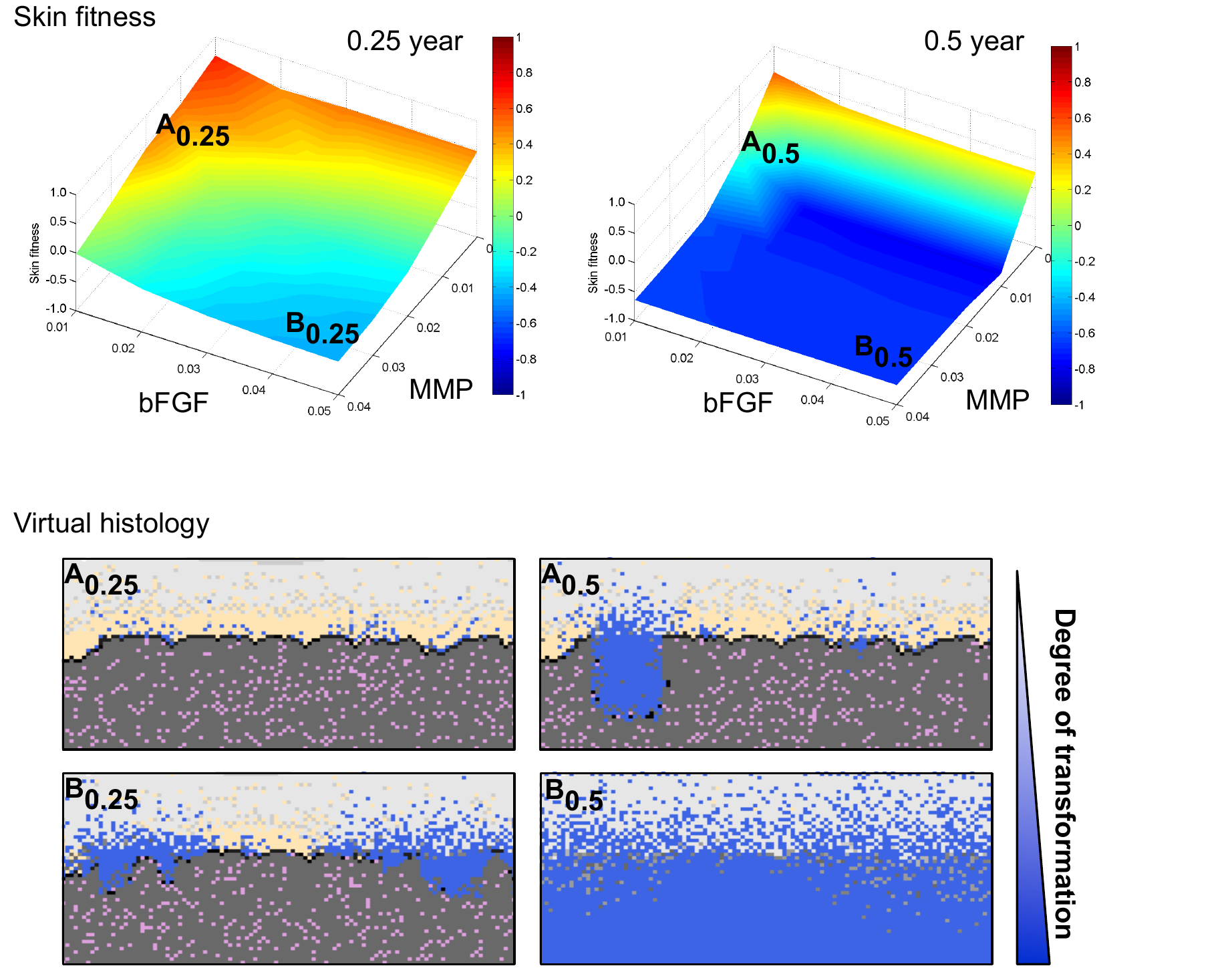}
\end{center}
\caption{
{\bf Degree of melanocyte malignancy and its effects on melanoma progression.} The degree of melanocyte transformation was varied by increasing production rate of bFGF and MMP. The parameter range was given as $(\beta_{b}^s,\beta_{P}^s) \in \{ (x,y)|  0.01 \leq x \leq 0.05, \  0.0 \leq y \leq 0.04 \}$, where $\beta_p^s$ represents bFGF production rate and $\beta_p^s$ describes MMP production rate of a transformed melanocyte. The first row shows the mean skin fitness landscape  after 0.25 and 0.5 years (averaged over 50 runs). $\mbox{A}_{0.25,0.5}$ represents the fitness from simulations with minimally transformed melanocytes (bFGF and MMP production rates of 0.01), while $\mbox{B}_{0.25,0.5}$ indicates the fitness achieved with more aggressive melanoma cells (bFGF production was 0.05 and MMP was 0.04. Lower panels show representative vSkin histology for each of these points (see figure \ref{Fig8} for a key of color coded cell types).}
\label{Fig10}
\end{figure}

\subsubsection*{Transformation of melanocytes drives melanoma initiation}
So far, our vSkin simulations have shown that senescent fibroblasts are ideal for disrupting the normal skin environment to enhance melanocyte proliferation and invasion. However, we know that melanocytes are significantly transformed in melanoma and so we also need to consider how such mutations will affect normal skin structure. We therefore integrated phenotypes of mutant melanocytes into the vSkin model in order to study the role of melanocyte transformation as a driver of melanoma initiation. Note that we assume all fibroblasts and keratinocytes are normal for these simulations. 

Just as we considered a spectrum of senescence in the fibroblast population, we examine a range of melanocyte transformation. Minimally transformed melanocytes have only lost growth inhibition and will proliferate even in the absence of bFGF.  As malignancy increases, melanocytes gain the ability to produce bFGF and MMP at ever increasing levels.  Therefore degree of malignancy specifically refers to how much bFGF and MMP the melanocytes produce, so the most malignant phenotype produces maximal bFGF and MMP. Interestingly, minimally mutated melanocytes (i.e. those that have only lost proliferative control) remained largely quiescent due to competition for space and growth factors as well as their inability to invade (results not shown). However, melanocytes with a greater mutational load proliferated rapidly and invaded taking over the dermal layer of the skin, as shown in Figure~\ref{Fig10}. and quantified by the change in fitness landscape at 0.25 years and 0.5 years. Since the more transformed melanocytes have autocrine signaling, they don't suffer from growth factor deficiency even with loss of fibroblasts. With any degree of malignancy, mutant melanocytes significantly lower skin fitness and do so in a very short time frame, contrast the 0.5 year landscape of Figure~\ref{Fig10} to the 6 year landscape of Figure~\ref{Fig9}. The effects become more dominant as the degree of transformation increases ($A_{0.25, 0.5} \rightarrow B_{0.25, 0.5}$). The mutant melanocytes disintegrate skin structure and completely take over as shown in a sample snapshot at $B_{0.5}$. This vSkin transformation recapitulates a type of melanoma which progresses very rapidly.  

\begin{figure}[!ht]
\begin{center}
\includegraphics[width = 3in]{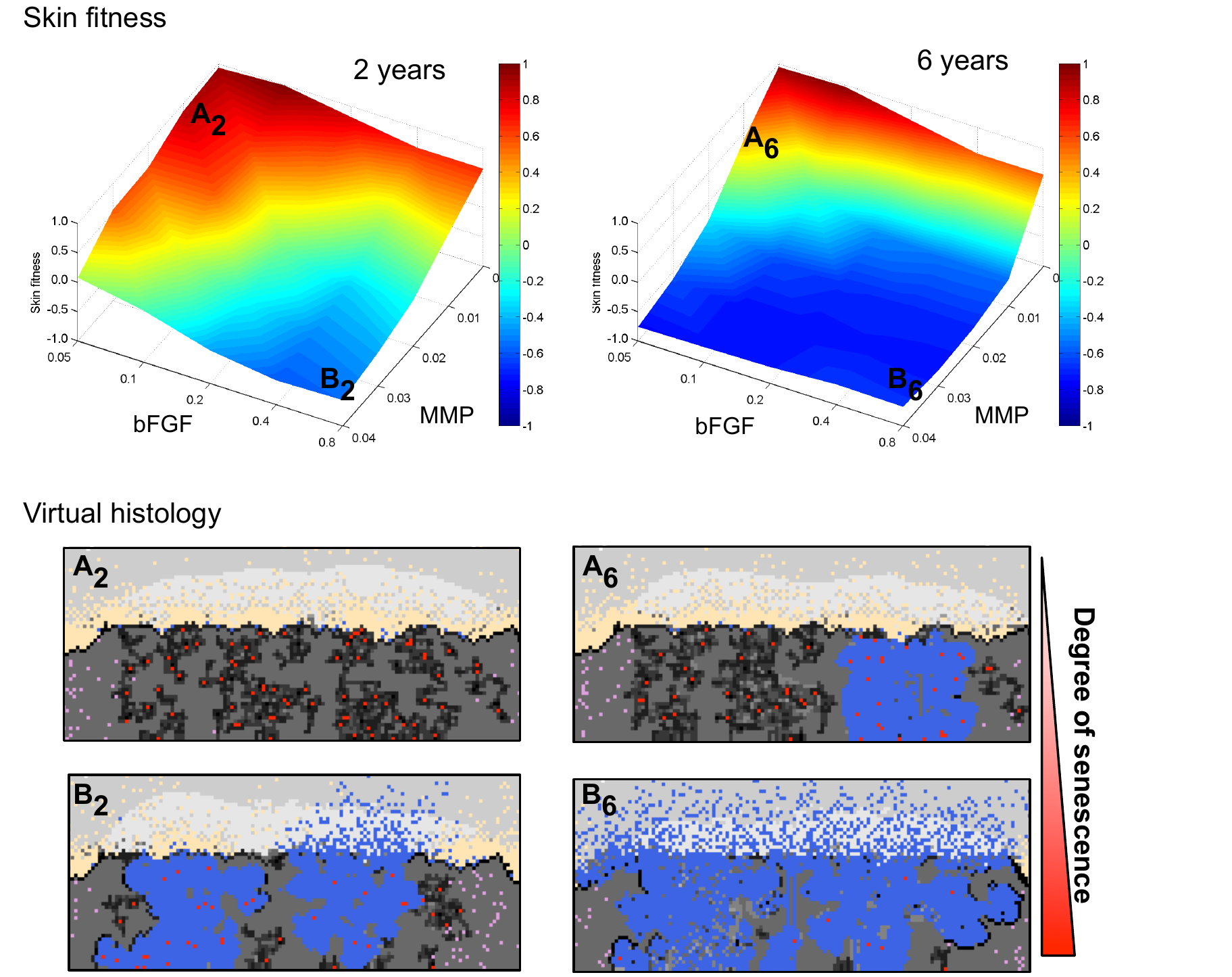}
\end{center}
\caption{
{\bf The synergistic effects of senescent fibroblasts and minimally transformed melanocytes in driving melanoma development.} The degree of senescence was varied as in the Figure~\ref{Fig9}. The transformed melanocytes (blue) are assumed to be only minimal transformed, i.e., loss of growth inhibition. Mildly senescent fibroblasts $\mbox{A}_{2,6}$ still manage to produce a small melanoma nodule (lower panel) but only impact skin fitness a little (upper panel). However, increasing the fibroblast senescence significantly impacts melanoma progression, destroying the skin (virtual histology, $\mbox{B}_{2,6}$) and reducing the fitness all the way to zero (upper panel). See figure \ref{Fig8} for a key of color coded cell types.}
\label{Fig11}
\end{figure}

\begin{figure}[!h]
\begin{center}
\includegraphics[width = 4in]{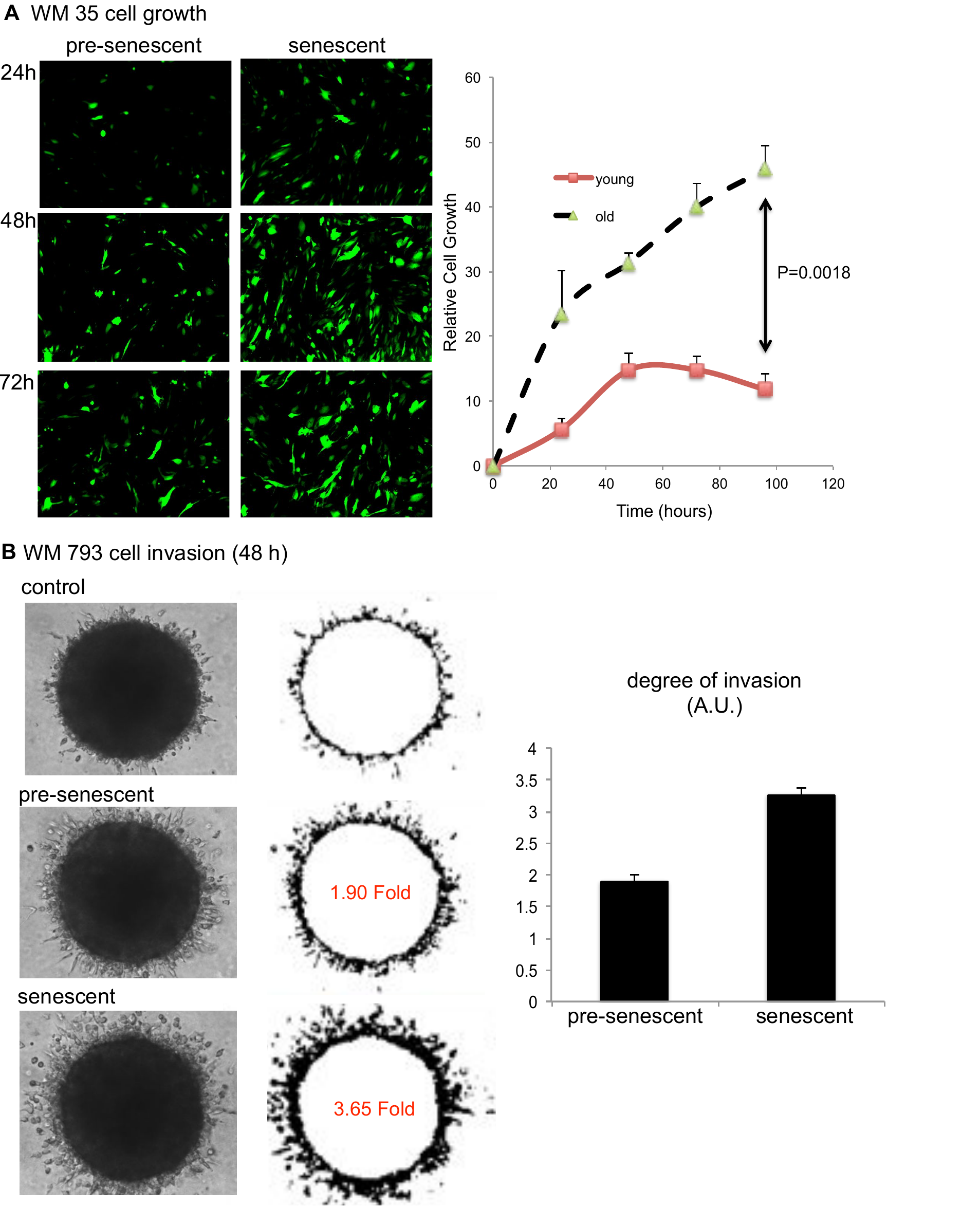}
\end{center}
\caption{ {\bf Senescent fibroblasts enhance melanoma cell growth and invasion.} A: Senescent fibroblasts aid the growth of early stage (non-tumorigenic) melanoma cells. Equal numbers of GFP-tagged (green) WM35 melanoma cells were seeded on top of either primary skin human fibroblasts (pre-senescent) or fibroblasts treated with radiation and allowed to undergo senescence (senescent) and allowed to grow for 72 hrs. Graph shows the mean (of three independent experiments) relative growth rates (relative to WM5 cells grown alone) where melanoma cells were plated on top of pre-senescent or senescent fibroblasts. B: Poorly invasive melanoma cells (WM793) were grown as spheroids alone, in co-culture with pre-senescent fibroblasts or in co-culture with senescent fibroblasts. The spheroids were then implanted in collagen and allowed to invade. Mean increases in invasion were calculated using ImageJ.}
\label{expSum}
\end{figure}

\subsubsection*{Transformed melanocytes exploit senescent fibroblasts to drive melanoma progression}
The next logical step was to integrate both transformed melanocytes and senescent fibroblasts together in the vSkin model, to investigate how interactions  between these abnormal populations drive melanoma initiation and progression. In order to highlight the impact of these interactions we only considered minimally transformed melanocytes in the following simulation. It is worth noting that these minimally transformed melanocytes have only lost control over proliferation and are incapable of producing cancer in the vSkin model independently since subsequent transformation would be needed for that to occur (i.e. resulting in increased bFGF and MMP production). 

Figure~\ref{Fig11} highlights how senescent fibroblasts help transformed melanocytes proliferate more rapidly and invade into the dermis. Even when minimally transformed melanocytes are combined with the least senescent fibroblasts in vSkin we see a positive impact on the melanoma growth, with a nodule of melanocytes (Figure~\ref{Fig11}A ) forming after 6 years.  Increasing the degree of senescence in the fibroblast population results in a more striking effect with the transformed melanocytes taking over the dermal layer, especially those nearby senescent fibroblasts (Figure~\ref{Fig11}B). From these results we can conclude that strong secretory phenotypes (e.g. the fibroblasts in Figure~\ref{Fig11} $B_{2,6}$) can transform essentially normal skin (with the exception of the initial mutant melanocyte) to a pathologic state (see density graph $B_{2,6}$). Therefore, senescent fibroblasts have the potential to propagate and maintain melanoma development, provided a minimally transformed melanocyte population already exists.

\subsection*{{\textbf{\textit{In vitro}} }experiments show senescent fibroblasts enhance melanocyte/melanoma growth and invasion}

So far, we have presented vSkin model predictions that senescent fibroblasts change the skin microenvironment and produce an aberrant skin structure that results from the overgrowth and invasion of normal melanocytes and early stage melanoma cells. In this section, we compare the model predictions with an {\it{in vitro}} co-culture assay.

\begin{figure}[p]
\begin{center}
\includegraphics[width = 4in]{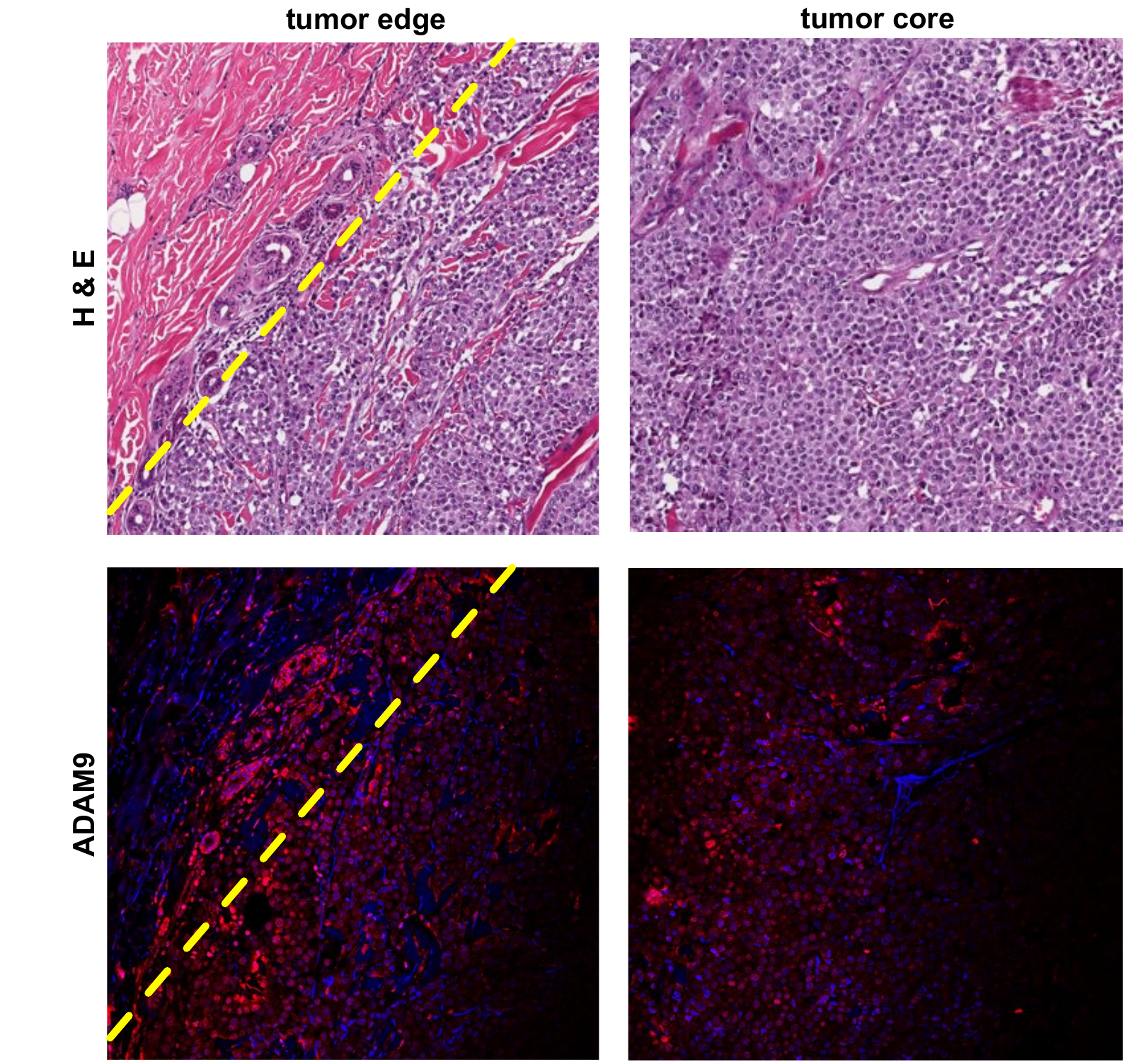}
\end{center}
\caption{
{\bf Melanoma tissue staining.} Representative H\&E-stained section of a nodule of metastatic melanoma (upper panel), stained with anti-ADAM9 (lower panel). The yellow line demarcates the tumor-stroma interface. There is high ADAM9 expression at the tumor/host interface in the melanoma specimen but weaker, more diffuse expression in the tumor core.}
\label{ExpFig2}
\end{figure}

We explored the prediction that senescent fibroblasts enhanced the growth and invasion of minimally transformed melanocytes in a series of {\it in vitro} fibroblast co-culture models. It was noted that plating early stage WM35 melanoma cells (which are non-tumorigenic in mice) on top of senescent fibroblasts enhanced their growth compared to non-senescent fibroblasts (Figure~\ref{expSum} A,B). To study invasion in a tissue-like context, we next generated spheroids in which poorly invasive melanoma cells (WM793) were co-cultured with either non-senescent or senescent human skin fibroblasts in a 3D collagen gel (Figure~\ref{expSum} C,D). It was noted that the WM793 cells invaded the collagen minimally when grown alone and that their invasive behavior was markedly increased when co-cultured with the senescent fibroblasts. Although not direct comparison with the vSkin results, our co-culture cell growth assay highlights one of predictions that the vSkin model makes, which is that senescent fibroblasts promote the growth of early stage melanoma cells. 
The potential clinical relevance of our model predictions and experimental findings was suggested by the observation that human melanoma specimens stained positively for the protease ADAM9 at the leading edge where the melanoma cells and fibroblasts interact (Figure~\ref{ExpFig2}). In contrast, little ADAM9 staining was noted in the tumor core where stromal infiltration was lacking.

\begin{figure}[p]
\begin{center}
\includegraphics[width = 4in]{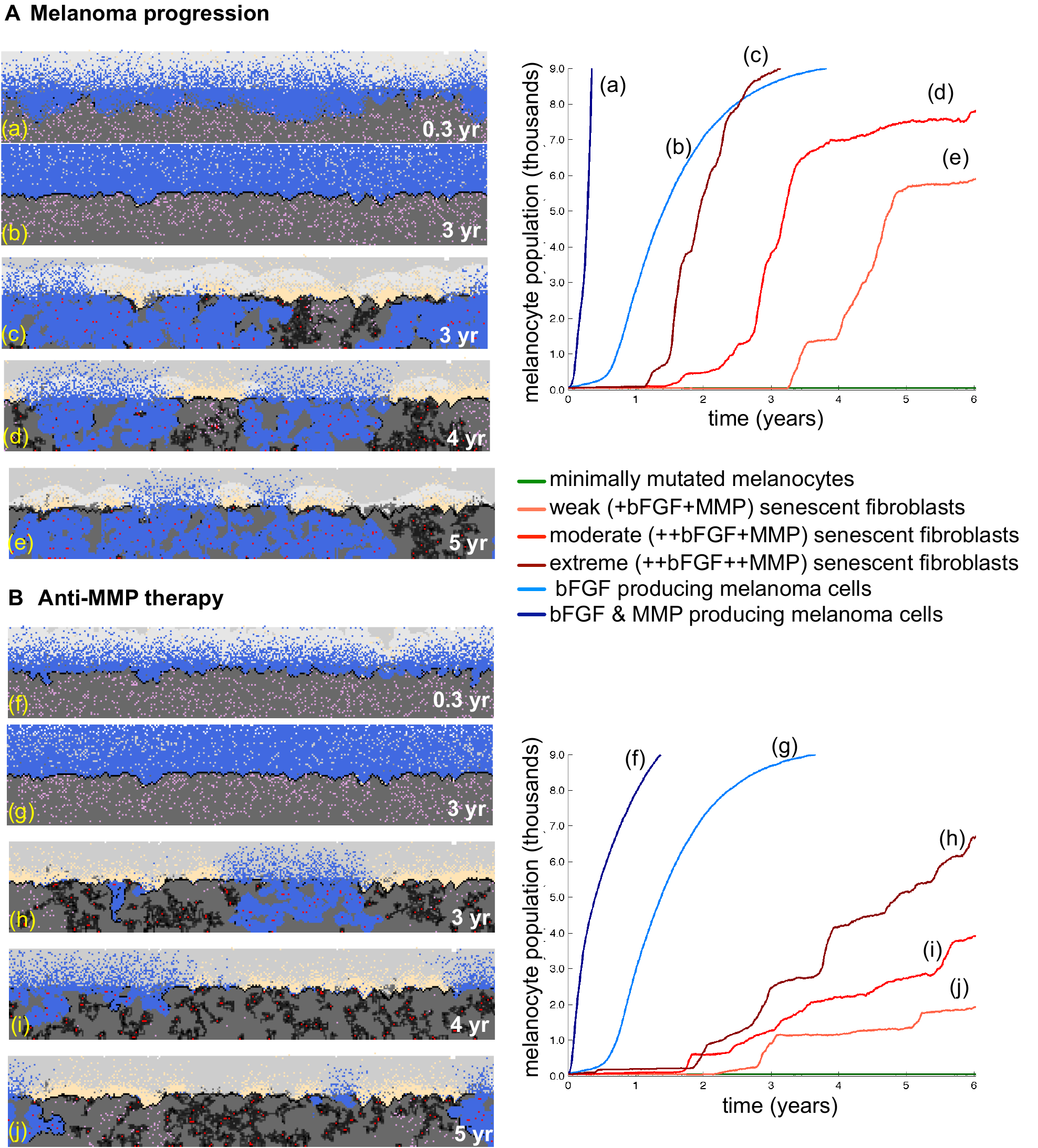}
\end{center}
\caption{
{\bf Multiple routes of Melanoma progression and a potential Anti-MMP therapy.} The total population of transformed melanocytes was quantified for all untreated (upper right) and Anti-MMP treated (bottom right) cases as a measure of tumor burden. Representative vSkin histology for each of these cases is also shown (left panels), see figure \ref{Fig8} for a key of color coded cell types.  A: The population of minimally mutated melanocytes (barely visible green line) remains constant. The most mutated melanocytes produce both bFGF and MMP and progress the most rapidly ((a), dark blue line). The melanocytes which produce bFGF alone, form melanoma {\it in situ} where the melanoma cells are constrained to the epidermis ((b), light blue line). The most senescent fibroblasts ((c), dark red line) help minimally mutated melanocytes to proliferate rapidly and progress to full melanoma. Senescent fibroblasts with weaker secretory phenotypes lead to a slower melanocyte proliferation and invaaion ((d), red line and (e), orange line). B: Melanoma progression after anti-MMP treatment. Anti-MMP therapy effects the growth of all populations but is more effective for melanoma progression driven by senescent fibroblasts ((h)-(j)) than melanocyte transformation driven melanomas ((f) and (g)). }
\label{Fig12}
\end{figure}

\section*{Discussion}

The regulation of skin homeostasis involves a complex interplay between different cell types and their microenvironment. To better understand how these interactions produce skin homeostasis, we developed a mathematical model that incorporates the key cellular (keratinocytes, melanocytes, and fibroblasts) and microenvironmental ($\mbox{TGF}_\beta$, EGF, bFGF and ECM) components of normal skin.  Our virtual skin model (vSkin) successfully recapitulated normal skin structure, skin homeostasis and wound healing responses. Surprisingly the homeostasis that emerged from this model was remarkably robust, being able to recover from both physical and biochemical perturbations. However, the recovery was altered  by both the perturbation type and period. With short-term perturbations inducing increased population growth, followed by a return to stable homeostasis. More persistent microenvironmental perturbations (specifically with bFGF) led to an altered but stable homeostatic state characterized by an increased level of melanocytes. One key perturbation that the system wasn't robust to was over-production of MMP and even then it was only when MMP levels were elevated for significant periods of time. Although, this isn't too surprising as MMPs are not normally produced in normal skin. These results indicate the importance of microenvironmental factors in maintaining skin homeostasis and highlight both bFGF and MMP as key factors in disrupting this homeostasis. 

The dialogue between our three cell types, keratinocytes, melanocytes, and fibroblasts is primarily mediated by growth factors produced by fibroblasts and cell-cell contact. One of the primary reasons for incorporating fibroblasts into vSkin was to understand their role in normal skin homeostasis, disruption and cancer initiation. An important experimental observation that motivated this line of inquiry was that when fibroblasts take on a secretory phenotype (such as that seen when undergoing senescence) they produce increased levels of bFGF and matrix degrading proteases (Figure~\ref{ExpFig1}), the very factors that normal skin seems to be most sensitive to. By introducing a population of senescent fibroblasts into vSkin we made the unexpected prediction that such secretory cells may play a central role in the initiation and progression of melanoma. The least senescent fibroblasts caused a temporary disruption of homeostasis but as we increased the level of senescence (i.e. increasing the levels of bFGF and MMP production) we were able to permanently disrupt the skin structure and produced melanocytic hyperplasia i.e. a mole like structure. However, if the senescent fibroblasts were to be removed, normal skin homeostasis would be reintroduced since the melanocytes are normal and only reacting to the altered microenvironment created by the secretory cells. 

Our vSkin model is the first model showing all the steps in skin carcinogenesis, from normal skin followed by melanocytic hyperplasia, and ending with melanoma. The activation of stroma due to senescence induces excessive growth of normal melanocytes, resulting in the development of melanocytic hyperplasia (Figure \ref{Fig9}). The vSkin model also recapitulated melanoma development driven only by melanocyte transformation (Figure \ref{Fig10}). In addition, the vSkin model proposes alternative routes to melanoma initiation, by showing that melanoma can occur as a result of co-operation between mutant melanocytes and senescent fibroblasts (Figure \ref{Fig11}). Given that most known driver mutations in melanoma are associated with entry into oncogene-induced senescence (OIS), our model suggests the possibility that signals from the senescent stroma may aid melanocyte transformation by allowing them to exit OIS. There is already good evidence that increased  PI3K/AKT signaling can overcome {\it BRAF} V600E-induced OIS in melanocytes~\cite{Vredeveld:2012ys} and significantly most of the growth factors and altered ECM interactions we describe here are known to exert their effects through the PI3K/AKT pathway~\cite{Dey:2010fk}.

Biochemical analysis shows that the alterations in senescent fibroblasts included increased expression of multiple growth factors and proteases. These alterations then stimulated the growth of normal melanocytes and early stage melanoma cells. The increased level of protease in senescent fibroblasts enhanced the invasiveness of late stage melanoma cells. Among many proteases, ADAM9 was found to be a putative mediator of matrix remodeling in senescent dermal fibroblasts. Importantly, the analysis of ADAM9 expression in melanoma samples from patients revealed that the expression was increased at the host-tumor interface, but not within the tumor core (Figure~\ref{ExpFig2}). Our findings mirror previous studies implicating a role for ADAM9 expression in melanoma cell/fibroblast interactions at the invasive front that is required for MMP-1 and MMP-2 mediated matrix degradation~\cite{Zigrino:2011ly}. Whilst not conclusive, the ADAM9 staining implicates the stroma is producing matrix degrading enzymes. This clinical observation along with all of the experimental results we presented were consistent with our model predictions that senescent fibroblasts contribute to melanoma initiation and early melanoma progression.

Our vSkin model proposes multiple routes to melanoma initiation and progression, summarized in the upper panel of Figure~\ref{Fig12}. The obvious and rapid route shown by the blue lines is due to the accumulation of mutations in melanocytes to drive melanoma development. The alternative route involves cooperation of mutant melanocytes with senescent fibroblasts, Figure~\ref{Fig12} (red lines). When senescent fibroblasts stimulate and maintain the excessive growth of minimally transformed melanocytes we observe the slowest melanoma growth. As senescent fibroblasts adopt stronger secretory phenotypes, melanoma progresses more rapidly.

If our predictions stand up to future validation, the fibroblasts could represent an important target for melanoma chemoprevention. The vSkin transformation to melanoma suggests a possible novel treatment, by negating the influence malignant phenotypes it may be possible to normalize the skin microenvironment and return (or maintain) skin homeostasis. Since proteases are one of key contributing factors driving melanoma initiation and progression, both from the senescent fibroblast and transformed melanocyte perspective, anti-MMP therapy might be a good treatment option. Indeed, there is already evidence that mutations in MMP-8, leading to an inactivation of protease activity that shifts the pattern of matrix degradation and growth factor availability, is an important event in the genesis of $\sim$23\% of all melanomas~\cite{Palavalli:2009uq}. As a purely hypothetical exercise we ran a suite of simulations to test this idea, almost as if we applied anti-MMP cream to the vSkin model such that MMP expression was blocked completely. Specifically, the MMP level is completely knocked down to zero uniformly in the whole domain, when the size of initial melanocyte population had doubled. This knock down is maintained until the simulation is finished. Our {\it in silico} anti-MMP treatment experiments show that the treatment was far more effective for the stroma dependent melanoma in that the population growth rates were significantly decreased (see  Figure~\ref{Fig12} (lower panel)).

This study emphasizes the importance of normal skin homeostasis and its disruption in melanoma initiation and progression. The possible role of MMP inhibition as a chemoprevention strategy for melanoma will require extensive future experimental validation. MMPs and other related proteases constitute a large family of enzymes with distinct pro and anti-oncogenic activities. The effective use of MMP blocking drugs is likely to require an exquisite sensitivity profile that current inhibitors do not yet offer. Our integrated approach highlighted two key unifying factors in disrupting 
skin homeostasis, MMP and bFGF production. Whether these factors were produced by transformed melanocytes or senescent fibroblasts led to similar skin degeneration. The most important implication is that we cannot only consider melanoma as a transformation of the melanocyte population but must also consider the fibroblast population. It is perhaps no accident then that melanoma really is a disease of the aged~\cite{Chao:2004kx,Lachiewicz:2008uq,Howlader2012}, where fibroblast senescence may naturally occur as we age and aid minor melanocyte mutations (due to a lifetime of skin sun exposure) to become full blown melanoma.

\section*{Acknowledgments}
We thank Dr. David Basanta for his initial input on the cell interaction network and crucially for providing his code, used in~\cite{Basanta:2009p10235}, that was used as the foundation for the vSkin code.

\section*{Author Contributions}

 Conceived and designed the computational model: EK and AA. Contributed initial implementation of the computational model: DB. Performed computational simulations: EK. Conceived and designed the experiments: KS. Performed the experiments: VR and IF. Analyzed the data: EK, JM, KS and AA.  Contributed reagents/materials: JM, RM, and SE. Wrote the paper: EK, KS and AA.

\bibliography{mynormalskin.bib}

\begin{thebibliography}{10}
\providecommand{\url}[1]{\texttt{#1}}
\providecommand{\urlprefix}{URL }
\expandafter\ifx\csname urlstyle\endcsname\relax
  \providecommand{\doi}[1]{doi:\discretionary{}{}{}#1}\else
  \providecommand{\doi}{doi:\discretionary{}{}{}\begingroup
  \urlstyle{rm}\Url}\fi
\providecommand{\bibAnnoteFile}[1]{%
  \IfFileExists{#1}{\begin{quotation}\noindent\textsc{Key:} #1\\
  \textsc{Annotation:}\ \input{#1}\end{quotation}}{}}
\providecommand{\bibAnnote}[2]{%
  \begin{quotation}\noindent\textsc{Key:} #1\\
  \textsc{Annotation:}\ #2\end{quotation}}
\providecommand{\eprint}[2][]{\url{#2}}

\bibitem{Rook:2010fk}
Rook A, Burns T (2010) Rook's textbook of dermatology.
\newblock Chichester, West Sussex, UK: Wiley-Blackwell, 8th edition.
\bibAnnoteFile{Rook:2010fk}

\bibitem{FITZPATRICK:1963fk}
Fitzpatrick TB, Breathnach AS (1963) The epidermal melanin unit system.
\newblock Dermatol Wochenschr 147: 481-9.
\bibAnnoteFile{FITZPATRICK:1963fk}

\bibitem{Haass:2005p106}
Haass N, Herlyn M (2005) Normal human melanocyte homeostasis as a paradigm for
  understanding melanoma.
\newblock Journal of Investigative Dermatology Symposium 10: 153-63.
\bibAnnoteFile{Haass:2005p106}

\bibitem{Costin:2007p2465}
Costin G, Hearing V (2007) Human skin pigmentation: melanocytes modulate skin
  color in response to stress.
\newblock FASEB J 21: 976-94.
\bibAnnoteFile{Costin:2007p2465}

\bibitem{Sorrell:2004bh}
Sorrell JM, Caplan AI (2004) Fibroblast heterogeneity: more than skin deep.
\newblock J Cell Sci 117: 667-75.
\bibAnnoteFile{Sorrell:2004bh}

\bibitem{Kalluri:2006uq}
Kalluri R, Zeisberg M (2006) Fibroblasts in cancer.
\newblock Nat Rev Cancer 6: 392-401.
\bibAnnoteFile{Kalluri:2006uq}

\bibitem{Chin:2003uq}
Chin L (2003) The genetics of malignant melanoma: lessons from mouse and man.
\newblock Nat Rev Cancer 3: 559-70.
\bibAnnoteFile{Chin:2003uq}

\bibitem{Miller:2006fk}
Miller AJ, Mihm MC Jr (2006) Melanoma.
\newblock N Engl J Med 355: 51-65.
\bibAnnoteFile{Miller:2006fk}

\bibitem{Albino:1989kx}
Albino AP, Nanus DM, Mentle IR, Cordon-Cardo C, McNutt NS, et~al. (1989)
  Analysis of ras oncogenes in malignant melanoma and precursor lesions:
  correlation of point mutations with differentiation phenotype.
\newblock Oncogene 4: 1363-74.
\bibAnnoteFile{Albino:1989kx}

\bibitem{Davies:2002fk}
Davies H, Bignell GR, Cox C, Stephens P, Edkins S, et~al. (2002) Mutations of
  the \emph{{B}{R}{A}{F}} gene in human cancer.
\newblock Nature 417: 949-54.
\bibAnnoteFile{Davies:2002fk}

\bibitem{Omholt:2003vn}
Omholt K, Platz A, Kanter L, Ringborg U, Hansson J (2003) \emph{{N}{R}{A}{S}}
  and \emph{BRAF} mutations arise early during melanoma pathogenesis and are
  preserved throughout tumor progression.
\newblock Clin Cancer Res 9: 6483-8.
\bibAnnoteFile{Omholt:2003vn}

\bibitem{Houben:2008uq}
Houben R, Vetter-Kauczok CS, Ortmann S, Rapp UR, Broecker EB, et~al. (2008)
  Phospho-{E}{R}{K} staining is a poor indicator of the mutational status of
  \emph{{B}{R}{A}{F}} and \emph{{N}{R}{A}{S}} in human melanoma.
\newblock J Invest Dermatol 128: 2003-12.
\bibAnnoteFile{Houben:2008uq}

\bibitem{Pollock:2003ys}
Pollock PM, Harper UL, Hansen KS, Yudt LM, Stark M, et~al. (2003) High
  frequency of \emph{{B}{R}{A}{F}} mutations in nevi.
\newblock Nat Genet 33: 19-20.
\bibAnnoteFile{Pollock:2003ys}

\bibitem{Michaloglou:2005p9307}
Michaloglou C, Vredeveld LCW, Soengas MS, Denoyelle C, Kuilman T, et~al. (2005)
  \emph{{B}{R}{A}{F}} {V}600{E}-associated senescence-like cell cycle arrest of
  human naevi.
\newblock Nature 436: 720--4.
\bibAnnoteFile{Michaloglou:2005p9307}

\bibitem{Dhomen:2009vn}
Dhomen N, Reis-Filho JS, da~Rocha~Dias S, Hayward R, Savage K, et~al. (2009)
  Oncogenic \emph{{BRAF}} induces melanocyte senescence and melanoma in mice.
\newblock Cancer Cell 15: 294-303.
\bibAnnoteFile{Dhomen:2009vn}

\bibitem{Vredeveld:2012ys}
Vredeveld LCW, Possik PA, Smit MA, Meissl K, Michaloglou C, et~al. (2012)
  Abrogation of \emph{BRAF} {V}600{E}-induced senescence by {PI}3{K} pathway
  activation contributes to melanomagenesis.
\newblock Genes Dev 26: 1055-69.
\bibAnnoteFile{Vredeveld:2012ys}

\bibitem{Krtolica:2001zr}
Krtolica A, Parrinello S, Lockett S, Desprez PY, Campisi J (2001) Senescent
  fibroblasts promote epithelial cell growth and tumorigenesis: a link between
  cancer and aging.
\newblock Proc Natl Acad Sci U S A 98: 12072-7.
\bibAnnoteFile{Krtolica:2001zr}

\bibitem{Krtolica:2002ly}
Krtolica A, Campisi J (2002) Cancer and aging: a model for the cancer promoting
  effects of the aging stroma.
\newblock Int J Biochem Cell Biol 34: 1401-14.
\bibAnnoteFile{Krtolica:2002ly}

\bibitem{Bhowmick:2004p7596}
Bhowmick N, Neilson E, Moses H (2004) Stromal fibroblasts in cancer initiation
  and progression.
\newblock Nature 18: 332-7.
\bibAnnoteFile{Bhowmick:2004p7596}

\bibitem{Parrinello:2005ve}
Parrinello S, Coppe JP, Krtolica A, Campisi J (2005) Stromal-epithelial
  interactions in aging and cancer: senescent fibroblasts alter epithelial cell
  differentiation.
\newblock J Cell Sci 118: 485-96.
\bibAnnoteFile{Parrinello:2005ve}

\bibitem{Coppe:2008p7159}
Copp{\'e} J, Patil C, Rodier F, Sun Y, Mu{\~n}oz D (2008) Senescence-associated
  secretory phenotypes reveal cell-nonautonomous functions of oncogenic
  \emph{{RAS}} and the p53 tumor suppressor.
\newblock PLoS Biol 6: 2853-68.
\bibAnnoteFile{Coppe:2008p7159}

\bibitem{Laberge:2012kh}
Laberge RM, Awad P, Campisi J, Desprez PY (2012) Epithelial-mesenchymal
  transition induced by senescent fibroblasts.
\newblock Cancer Microenviron 5: 39-44.
\bibAnnoteFile{Laberge:2012kh}

\bibitem{Campisi:1998fu}
Campisi J (1998) The role of cellular senescence in skin aging.
\newblock J Investig Dermatol Symp Proc 3: 1-5.
\bibAnnoteFile{Campisi:1998fu}

\bibitem{Ressler:2006zr}
Ressler S, Bartkova J, Niederegger H, Bartek J, Scharffetter-Kochanek K, et~al.
  (2006) p16{INK}4{A} is a robust in vivo biomarker of cellular aging in human
  skin.
\newblock Aging Cell 5: 379-89.
\bibAnnoteFile{Ressler:2006zr}

\bibitem{Zglinicki:2000qc}
von Zglinicki T (2000) Role of oxidative stress in telomere length regulation
  and replicative senescence.
\newblock Ann N Y Acad Sci 908: 99-110.
\bibAnnoteFile{Zglinicki:2000qc}

\bibitem{Zglinicki:2002ss}
von Zglinicki T (2002) Oxidative stress shortens telomeres.
\newblock Trends Biochem Sci 27: 339-44.
\bibAnnoteFile{Zglinicki:2002ss}

\bibitem{Bedogni:2005mw}
Bedogni B, Welford SM, Cassarino DS, Nickoloff BJ, Giaccia AJ, et~al. (2005)
  The hypoxic microenvironment of the skin contributes to {AKT}-mediated
  melanocyte transformation.
\newblock Cancer Cell 8: 443-54.
\bibAnnoteFile{Bedogni:2005mw}

\bibitem{Li:2003p77}
Li G, Satyamoorthy K, Meier F, Berking C (2003) Function and regulation of
  melanoma-stromal fibroblast interactions: when seeds meet soil.
\newblock Oncogene 22: 3162-71.
\bibAnnoteFile{Li:2003p77}

\bibitem{Valeyev:2010fk}
Valeyev NV, Hundhausen C, Umezawa Y, Kotov NV, Williams G, et~al. (2010) A
  systems model for immune cell interactions unravels the mechanism of
  inflammation in human skin.
\newblock PLoS Comput Biol 6: e1001024.
\bibAnnoteFile{Valeyev:2010fk}

\bibitem{Aylaj:2011uq}
Aylaj B, Luciani F, Delmas V, Larue L, De~Vuyst F (2011) Melanoblast
  proliferation dynamics during mouse embryonic development. {M}odeling and
  validation.
\newblock J Theor Biol 276: 86-98.
\bibAnnoteFile{Aylaj:2011uq}

\bibitem{Eikenberry:2009p2090}
Eikenberry S, Thalhauser C, Kuang Y (2009) Tumor-immune interaction, surgical
  treatment, and cancer recurrence in a mathematical model of melanoma.
\newblock Plos Comput Biol 5: e1000362.
\bibAnnoteFile{Eikenberry:2009p2090}

\bibitem{Basan:2011vn}
Basan M, Joanny JF, Prost J, Risler T (2011) Undulation instability of
  epithelial tissues.
\newblock Physical Review Letters 106: 158101.
\bibAnnoteFile{Basan:2011vn}

\bibitem{Ciarletta:2011kx}
Ciarletta P, Foret L, Ben~Amar M (2011) The radial growth phase of malignant
  melanoma: multi-phase modelling, numerical simulations and linear stability
  analysis.
\newblock Journal of the Royal Society Interface 8: 345--368.
\bibAnnoteFile{Ciarletta:2011kx}

\bibitem{Chatelain:2011ys}
Chatelain C, Balois T, Ciarletta P, Ben~Amar M (2011) Emergence of
  microstructural patterns in skin cancer: a phase separation analysis in a
  binary mixture.
\newblock New Journal of Physics 13: 115013.
\bibAnnoteFile{Chatelain:2011ys}

\bibitem{Grabe:2005p8047}
Grabe N, Neuber K (2005) A multicellular systems biology model predicts
  epidermal morphology, kinetics and ca2+ flow.
\newblock Bioinformatics 21: 3541--7.
\bibAnnoteFile{Grabe:2005p8047}

\bibitem{Suetterlin:2009ys}
Suetterlin T, Huber S, Dickhaus H, Grabe N (2009) Modeling multi-cellular
  behavior in epidermal tissue homeostasis via finite state machines in
  multi-agent systems.
\newblock Bioinformatics 25: 2057--2063.
\bibAnnoteFile{Suetterlin:2009ys}

\bibitem{Adra:2010p3575}
Adra S, Sun T, MacNeil S, Holcombe M, Smallwood R (2010) Development of a three
  dimensional multiscale computational model of the human epidermis.
\newblock PLoS ONE 5: e8511.
\bibAnnoteFile{Adra:2010p3575}

\bibitem{Schaller:2007uq}
Schaller G, Meyer-Hermann M (2007) A modelling approach towards epidermal
  homoeostasis control.
\newblock J Theor Biol 247: 554-73.
\bibAnnoteFile{Schaller:2007uq}

\bibitem{Grabe:2007p7951}
Grabe N, Neuber K (2007) Simulating psoriasis by altering transit amplifying
  cells.
\newblock Bioinformatics 23: 1309--12.
\bibAnnoteFile{Grabe:2007p7951}

\bibitem{Sun:2008p3760}
Sun T, McMinn P, Holcombe M, Smallwood R (2008) Agent based modelling helps in
  understanding the rules by which fibroblasts support keratinocyte colony
  formation.
\newblock PLoS ONE 3: e2129.
\bibAnnoteFile{Sun:2008p3760}

\bibitem{Thingnes:2012fk}
Thingnes J, Lavelle TJ, Hovig E, Omholt SW (2012) Understanding the melanocyte
  distribution in human epidermis: an agent-based computational model approach.
\newblock PLoS One 7: e40377.
\bibAnnoteFile{Thingnes:2012fk}

\bibitem{anderson1997}
Anderson ARA, Sleeman BD, Young IM, Griffiths BS (1997) Nematode movement along
  a chemical gradient in a structurally heterogeneous environment. 2. {T}heory.
\newblock Fundam Appl Nematol 20: 165-172.
\bibAnnoteFile{anderson1997}

\bibitem{Anderson:1998fk}
Anderson AR, Chaplain MA (1998) Continuous and discrete mathematical models of
  tumor-induced angiogenesis.
\newblock Bull Math Biol 60: 857-99.
\bibAnnoteFile{Anderson:1998fk}

\bibitem{AndersonInAlt:2003p11378}
Anderson A, Pitairn A (2003) Application of the hybrid discrete-continuum
  technique, in polymer and cell dynamics, eds. {W}. {A}lt, {M}. {C}haplain,
  {M}. {G}riebel, {J}. {L}enz.
\newblock Birkhauser : 261-279.
\bibAnnoteFile{AndersonInAlt:2003p11378}

\bibitem{Anderson:2005p10258}
Anderson ARA (2005) A hybrid mathematical model of solid tumour invasion: the
  importance of cell adhesion.
\newblock Math Med Biol 22: 163--86.
\bibAnnoteFile{Anderson:2005p10258}

\bibitem{Basanta:2009p10235}
Basanta D, Strand DW, Lukner RB, Franco OE, Cliffel DE, et~al. (2009) The role
  of transforming growth factor-beta-mediated tumor-stroma interactions in
  prostate cancer progression: an integrative approach.
\newblock Cancer research 69: 7111--20.
\bibAnnoteFile{Basanta:2009p10235}

\bibitem{Green:1985vn}
Green MR, Couchman JR (1985) Differences in human skin between the epidermal
  growth factor receptor distribution detected by {EGF} binding and monoclonal
  antibody recognition.
\newblock J Invest Dermatol 85: 239-45.
\bibAnnoteFile{Green:1985vn}

\bibitem{Nanney:1986ys}
Nanney LB, Stoscheck CM, Magid M, King LE Jr (1986) Altered [125{I}] epidermal
  growth factor binding and receptor distribution in psoriasis.
\newblock J Invest Dermatol 86: 260-5.
\bibAnnoteFile{Nanney:1986ys}

\bibitem{Krane:1991zr}
Krane JF, Murphy DP, Carter DM, Krueger JG (1991) Synergistic effects of
  epidermal growth factor ({E}{G}{F}) and insulin-like growth factor
  {I}/somatomedin {C} ({I}{G}{F}-{I}) on keratinocyte proliferation may be
  mediated by {I}{G}{F}-{I} transmodulation of the {E}{G}{F} receptor.
\newblock J Invest Dermatol 96: 419-24.
\bibAnnoteFile{Krane:1991zr}

\bibitem{Hannon:1994fk}
Hannon GJ, Beach D (1994) p15{I}{N}{K}4{B} is a potential effector of
  {T}{G}{F}-beta-induced cell cycle arrest.
\newblock Nature 371: 257-61.
\bibAnnoteFile{Hannon:1994fk}

\bibitem{Ewen:1996uq}
Ewen ME (1996) p53-dependent repression of cdk4 synthesis in transforming
  growth factor-beta-induced {G}1 cell cycle arrest.
\newblock J Lab Clin Med 128: 355-60.
\bibAnnoteFile{Ewen:1996uq}

\bibitem{Shih:1993zr}
Shih IM, Herlyn M (1993) Role of growth factors and their receptors in the
  development and progression of melanoma.
\newblock J Invest Dermatol 100: 196S-203S.
\bibAnnoteFile{Shih:1993zr}

\bibitem{Makino:2010ly}
Makino T, Jinnin M, Muchemwa FC, Fukushima S, Kogushi-Nishi H, et~al. (2010)
  Basic fibroblast growth factor stimulates the proliferation of human dermal
  fibroblasts via the {E}{R}{K}1/2 and {J}{N}{K} pathways.
\newblock Br J Dermatol 162: 717-23.
\bibAnnoteFile{Makino:2010ly}

\bibitem{Andriani:2003ve}
Andriani F, Margulis A, Lin N, Griffey S, Garlick JA (2003) Analysis of
  microenvironmental factors contributing to basement membrane assembly and
  normalized epidermal phenotype.
\newblock J Invest Dermatol 120: 923-31.
\bibAnnoteFile{Andriani:2003ve}

\bibitem{Haass:2005p80}
Haass N, Smalley K, Li L (2005) Adhesion, migration and communication in
  melanocytes and melanoma.
\newblock Pigment Cell Melanoma Res 18: 150-9.
\bibAnnoteFile{Haass:2005p80}

\bibitem{NikolasKHaassKerianSMSmalleyLingL:2005p949}
Nikolas K~Haass LL Kerian S M~Smalley, Herlyn M (2005) Adhesion, migration and
  commnication in melanocyte and melanoma.
\newblock Pigment Cell Res 18: 150-9.
\bibAnnoteFile{NikolasKHaassKerianSMSmalleyLingL:2005p949}

\bibitem{Morelli:1993fk}
Morelli JG, Yohn JJ, Zekman T, Norris DA (1993) Melanocyte movement in vitro:
  role of matrix proteins and integrin receptors.
\newblock J Invest Dermatol 101: 605-8.
\bibAnnoteFile{Morelli:1993fk}

\bibitem{Swabb:1974p10166}
Swabb EA, Wei J, Gullino PM (1974) Diffusion and convection in normal and
  neoplastic tissues.
\newblock Cancer research 34: 2814--22.
\bibAnnoteFile{Swabb:1974p10166}

\bibitem{molecularweight}
The {U}{C}{S}{D} {S}ignaling {G}ateway.
\newblock \urlprefix\url{http://www.signaling-gateway.org/}.
\bibAnnoteFile{molecularweight}

\bibitem{Partridge:1989vn}
Partridge M, Green MR, Langdon JD, Feldmann M (1989) Production of
  {T}{G}{F}-alpha and {T}{G}{F}-beta by cultured keratinocytes, skin and oral
  squamous cell carcinomas--potential autocrine regulation of normal and
  malignant epithelial cell proliferation.
\newblock Br J Cancer 60: 542-8.
\bibAnnoteFile{Partridge:1989vn}

\bibitem{Amjad:2007uq}
Amjad SB, Carachi R, Edward M (2007) Keratinocyte regulation of {TGF}-beta and
  connective tissue growth factor expression: a role in suppression of scar
  tissue formation.
\newblock Wound Repair Regen 15: 748-55.
\bibAnnoteFile{Amjad:2007uq}

\bibitem{Halaban:1988qf}
Halaban R, Langdon R, Birchall N, Cuono C, Baird A, et~al. (1988) Basic
  fibroblast growth factor from human keratinocytes is a natural mitogen for
  melanocytes.
\newblock J Cell Biol 107: 1611-9.
\bibAnnoteFile{Halaban:1988qf}

\bibitem{Kim:2010fk}
Kim Y, Friedman A (2010) Interaction of tumor with its micro-environment: A
  mathematical model.
\newblock Bulletin of Mathematical Biology 72: 1029--1068.
\bibAnnoteFile{Kim:2010fk}

\bibitem{Tavakkol:1999fk}
Tavakkol A, Varani J, Elder JT, Zouboulis CC (1999) Maintenance of human skin
  in organ culture: role for insulin-like growth factor-1 receptor and
  epidermal growth factor receptor.
\newblock Arch Dermatol Res 291: 643-51.
\bibAnnoteFile{Tavakkol:1999fk}

\bibitem{Smalley:2006fk}
Smalley KSM, Haass NK, Brafford PA, Lioni M, Flaherty KT, et~al. (2006)
  Multiple signaling pathways must be targeted to overcome drug resistance in
  cell lines derived from melanoma metastases.
\newblock Mol Cancer Ther 5: 1136-44.
\bibAnnoteFile{Smalley:2006fk}

\bibitem{Brohem:2011fk}
Brohem CA, Cardeal LBdS, Tiago M, Soengas MS, Barros SBdM, et~al. (2011)
  Artificial skin in perspective: concepts and applications.
\newblock Pigment Cell Melanoma Res 24: 35-50.
\bibAnnoteFile{Brohem:2011fk}

\bibitem{Zigrino:2011ly}
Zigrino P, Nischt R, Mauch C (2011) The disintegrin-like and cysteine-rich
  domains of {A}{D}{A}{M}-9 mediate interactions between melanoma cells and
  fibroblasts.
\newblock J Biol Chem 286: 6801-7.
\bibAnnoteFile{Zigrino:2011ly}

\bibitem{Dey:2010fk}
Dey JH, Bianchi F, Voshol J, Bonenfant D, Oakeley EJ, et~al. (2010) Targeting
  fibroblast growth factor receptors blocks {PI3K}/{AKT} signaling, induces
  apoptosis, and impairs mammary tumor outgrowth and metastasis.
\newblock Cancer Res 70: 4151-62.
\bibAnnoteFile{Dey:2010fk}

\bibitem{Palavalli:2009uq}
Palavalli LH, Prickett TD, Wunderlich JR, Wei X, Burrell AS, et~al. (2009)
  Analysis of the matrix metalloproteinase family reveals that {MMP}8 is often
  mutated in melanoma.
\newblock Nat Genet 41: 518-20.
\bibAnnoteFile{Palavalli:2009uq}

\bibitem{Chao:2004kx}
Chao C, Martin RCG 2nd, Ross MI, Reintgen DS, Edwards MJ, et~al. (2004)
  Correlation between prognostic factors and increasing age in melanoma.
\newblock Ann Surg Oncol 11: 259-64.
\bibAnnoteFile{Chao:2004kx}

\bibitem{Lachiewicz:2008uq}
Lachiewicz AM, Berwick M, Wiggins CL, Thomas NE (2008) Epidemiologic support
  for melanoma heterogeneity using the surveillance, epidemiology, and end
  results program.
\newblock J Invest Dermatol 128: 1340-2.
\bibAnnoteFile{Lachiewicz:2008uq}

\bibitem{Howlader2012}
Howlader N, Noone A, Krapcho M, Neyman N, Aminou R, et~al. (2011) {SEER} Cancer
  Statistics Review, 1975-2009 (Vintage 2009 Populations) based on November
  2011 {SEER} data submission, posted to the {SEER} web site, April 2012.
\newblock National Cancer Institute. Bethesda, MD.
\bibAnnoteFile{Howlader2012}

\end{thebibliography}
\end{document}